\documentclass[a4paper,10pt,romanappendices
]{IEEEtran}

\usepackage{commath,amsmath,graphicx,epstopdf,amsthm,amssymb,float,cite,multicol,enumitem,xcolor}

\usepackage{acronym}

\acrodef{MSE}[MSE]{mean-squared error}
\acrodef{RMSE}[RMSE]{root mean-squared-error}
\acrodef{MMSE}[MMSE]{{minimum mean-squared-error}}
\acrodef{SNR}[SNR]{signal-to-noise ratio}
\acrodef{iid}[i.i.d.]{independent and identically distributed}
\acrodef{CRB}[CRB]{Cram$\acute{\text{e}}$r-Rao bound}
\acrodef{BCRB}[BCRB]{Bayesian Cram$\acute{\text{e}}$r-Rao bound}
\acrodef{AT-BCRB}[AT-BCRB]{asymptotically tight Bayesian Cram$\acute{\text{e}}$r-Rao bound}
\acrodef{ZZB}[ZZB]{Ziv-Zakai bound}
\acrodef{PDF}[PDF]{probability density function}
\acrodef{w.r.t.}[w.r.t.]{with respect to}
\acrodef{WWB}[WWB]{Weiss-Weinstein bound}
\acrodef{MAP}[MAP]{maximum {\it{a-posteriori}} probability}
\acrodef{ML}[ML]{maximum likelihood}
\acrodef{DOA}[DOA]{direction-of-arrival}
\acrodef{WBCRB}[WBCRB]{weighted BCRB}
\acrodef{ECRB}[ECRB]{expected CRB}
\acrodef{FIM}[FIM]{Fisher information matrix}

\newcommand{\avec}{{\bf{a}}}

\newcommand{\dvec}{{\bf{d}}}
\newcommand{\evec}{{\bf{e}}}
\newcommand{\fvec}{{\bf{f}}}

\newcommand{\wvec}{{\bf{w}}}
\newcommand{\xvec}{{\bf{x}}}

\newcommand{\vvec}{{\bf{v}}}

\newcommand{\onevec}{{\bf{1}}}
\newcommand{\zerovec}{{\bf{0}}}

\newcommand{\psivec}{{\boldsymbol{\psi}}}
\newcommand{\xivec}{{\boldsymbol{\xi}}}

\newcommand{\Lambdamat}{{\bf{\Lambda}}}

\newcommand{\Cmat}{{\bf{C}}}
\newcommand{\Dmat}{{\bf{D}}}

\newcommand{\Fmat}{{\bf{F}}}
\newcommand{\Gmat}{{\bf{G}}}
\newcommand{\Hmat}{{\bf{H}}}
\newcommand{\Jmat}{{\bf{J}}}

\newcommand{\Imat}{{\bf{I}}}
\newcommand{\Lmat}{{\bf{L}}}

\newcommand{\Smat}{{\bf{S}}}

\newcommand{\Umat}{{\bf{U}}}

\newcommand{\Wmat}{{\bf{W}}}

\newcommand{\Zmat}{{\bf{Z}}}

\newcommand{\define}{\triangleq}

\newcommand{\Psimat}{\mbox{\boldmath $\Psi$}}
\newcommand{\Phimat}{\mbox{\boldmath $\Phi$}}

\def\psivec{{\mbox{\boldmath $\psi$}}}

\def\varphivec{{\mbox{\boldmath $\varphi$}}}

\def\thetavec{{\boldsymbol{\theta}}}

\newcommand{\E}{{\rm{E}}}

\newcommand{\be}{\begin{equation}}
\newcommand{\ee}{\end{equation}}
\newcommand{\beqna}{\begin{eqnarray}}
\newcommand{\eeqna}{\end{eqnarray}}

\makeatletter
\def\user@resume{resume}
\def\user@intermezzo{intermezzo}
\newcounter{previousequation}
\newcounter{lastsubequation}
\newcounter{savedparentequation}
\makeatother 
\newcommand{\diff}{{\textnormal{d}}}

\newtheorem{Thm}{Theorem}

\newtheorem*{Proof}{Proof}

\newtheorem{Prop}[Thm]{Proposition}
\newtheorem{Lemm}[Thm]{Lemma}

\DeclareMathAlphabet\mathbfcal{OMS}{cmsy}{b}{n}

\title{Asymptotically Tight Bayesian Cram\'{e}r-Rao Bound}
\author{
Ori Aharon and Joseph Tabrikian,~\IEEEmembership{Fellow,~IEEE}
 \vspace{-20pt}
\thanks{{This research was partially supported by THE ISRAEL SCIENCE FOUNDATION (grant No. 2666/19 and 2493/23).\newline
\indent The authors are with the School of Electrical and Computer Engineering,
Ben-Gurion University of the Negev, Beer-Sheva 84105, Israel (e-mail: aharonor@post.bgu.ac.il; joseph@bgu.ac.il)
}
}

}
\begin{document}

\maketitle
\nopagebreak

\begin{abstract}
Performance bounds for parameter estimation play a crucial role in statistical signal processing theory and applications. Two widely recognized bounds are the Cram\'{e}r-Rao bound (CRB) in the non-Bayesian framework, and the Bayesian CRB (BCRB) in the Bayesian framework. However, unlike the CRB, the BCRB is asymptotically unattainable in general, and its equality condition is restrictive. This paper introduces an extension of the Bobrovsky--Mayer-Wolf--Zakai class of bounds, also known as the weighted BCRB (WBCRB). The WBCRB is optimized by tuning the weighting function in the scalar case. Based on this result, we propose an asymptotically tight version of the bound called AT-BCRB. We prove that the AT-BCRB is  asymptotically attained by the maximum {\it a-posteriori} probability (MAP) estimator. Furthermore, we extend the WBCRB and the AT-BCRB to the case of vector parameters. The proposed bounds are evaluated in several fundamental signal processing examples, such as variance estimation of white Gaussian process, direction-of-arrival estimation, and mean estimation of Gaussian process with unknown variance and prior statistical information. It is shown that unlike the BCRB, the proposed bounds are asymptotically attainable and coincide with the expected CRB (ECRB). The ECRB, which imposes uniformly unbiasedness, cannot serve as a valid lower bound in the Bayesian framework, while the proposed bounds are valid for any estimator.
\end{abstract}

\section{Introduction}
\label{sec:NBE:intro}
Performance bounds are fundamental in estimation theory and statistical signal processing, serving as benchmarks for performance evaluation of estimators and playing a crucial role in system design and feasibility study (see e.g. \cite{Tabrikian_Krolik,Athley_TAES,Athley_TSP,Moses_array_design,Li_Tabrikian_Nehorai,Keskin_Koivunen}). They are also employed as an optimization criterion in sequential cognitive systems 
\cite{Wasim_TSP,Chavali_Nehorai,Haykin_Xue_Setoodeh_Cognitive,Greco_Gini_Bell,Bell_cognitive_2015,Rubinstein_Tabrikian,Sharaga,Tabrikian_Antenna_Slection},
The most common criterion considered for performance bounds for parameter estimation is the \ac{MSE}. 

Key attributes of performance bounds include tightness, simplicity, and the ability to provide insight into the considered problem at hand. However, existing \ac{MSE} tight lower bounds in both Bayesian and non-Bayesian theories are often computationally challenging and lack analytical tools for system design and optimization. The \ac{CRB} \cite{CRB,Rao} and the \ac{BCRB} \cite{BCRB} are the most popular \ac{MSE} bounds in the non-Bayesian and the Bayesian approaches, respectively. They are favored for their computational efficiency and for their ability to provide insight into the problem under consideration.

An important property of the CRB is that under mild regularity conditions, it is asymptotically attainable by the maximum likelihood (ML) estimator. In contrast, the BCRB does not satisfy this property when the Fisher information of the conditional problem depends on the parameter. In fact, the BCRB is attainable only when the posterior distribution of the parameter is Gaussian. Surprisingly, currently there is no simple Bayesian bound that is known to be asymptotically attainable, and this intriguing issue remains an open research challenge.  

Numerous efforts have been dedicated to obtain tight MSE bounds in the Bayesian framework. These Bayesian bounds can be categorized into two main families: 1) the Ziv-Zakai family which relies on Chebyshev's inequality, includes e.g. \cite{Ziv_Zakai,Bellini_Tartara,Bell_Ziv_Zakai,Basu_Bresler}, and 2) the Weiss-Weinstein family, which is based on the Cauchy-Schwarz inequality, including works like  \cite{WWB,WW_Class,RMB,BZB,renaux2008fresh,Todros_Tabrikian2,Chaumette_Renaux_ElKorso,aharon_tabrikian_WWTB}. Another class of Bayesian lower bounds was proposed in \cite{BenHaim_Eldar_OBB} based on non-Bayesian bounds for biased estimators, while optimizing the bias function. These bounds primarily aim to predict the threshold phenomenon.  The popularity of these bounds has been limited due to their computational complexity and the fact that they do not provide insight into the underlying problem. A recent contribution within the Weiss-Weinstein family, presented in  \cite{aharon_tabrikian_WWTB}, introduced a computationally efficient bound, which is directly related to the ambiguity function, providing insight into the considered problem. However, the general asymptotic tightness of this newly proposed bound remains unguaranteed. 

Since the introduction of the BCRB, significant efforts have been made in order to derive tighter Cram\'{e}r-Rao-type bounds. A general class of bounds was proposed in \cite{BMZ} called Bobrovsky--Mayer-Wolf--Zakai bound (BMZB), which generalizes the BCRB. However, it involves selection of a weighting function, which remains a challenge. In \cite{Miller_Chang}, Miller and Chang proposed a Bayesian bound which is tighter than the BCRB,  but it assumes unbiasedness in the strict sense for part of the unknown parameters, and thus, it cannot serve as a lower bound on the \ac{MSE} of any estimator. In \cite{tabrikian_krolik_TAP,Trees} the asymptotic performance of the \ac{MAP} estimator was studied and shown to be equal to the \ac{ECRB}. It can be shown that the \ac{ECRB} can serve as a lower bound for any uniformly unbiased estimator, and thus, similar to \cite{Miller_Chang}, it cannot serve as a valid Bayesian lower bound for any estimator. In \cite{Tight_BCRB_Frische_Chamette_FUSION,Tight_BCRB_Frische_Chamette} a tight version of the BCRB was proposed, but its computation is difficult since it involves expectations \ac{w.r.t.} observations, similar to calculation of the \ac{MMSE} estimator. The BCRB was adapted to cover various cases, such as, presence of nuisance deterministic parameters \cite{PartI} prior distribution with bounded support \cite{Scaglione_Bounded_Support}, periodic problems \cite{BPCRB}. 

In this paper, we propose a simple and asymptotically tight version of the BCRB, called \ac{AT-BCRB}. The scalar version of the proposed bound is related to the BMZB, in which the weighting is chosen to provide a simple and asymptotically tight bound, which is attained by the \ac{ML} or the \ac{MAP} estimators. The proposed bound is obtained using the Cauchy-Schwarz inequality, with modification of the auxiliary function of the BCRB. 

The main contributions of this paper are: 
\begin{itemize}
    \item Derivation of the \ac{WBCRB} as a function of the conditional Fisher information matrix (FIM) for scalar and vector parameters. 
\item Numerical optimization of the proposed \ac{WBCRB} \ac{w.r.t.} the weighting function. 
    \item  An asymptotically tight BCRB, called \ac{AT-BCRB}, is proposed and presented in a closed-form for scalar and vector parameters.
    \item The relationships between the proposed \ac{AT-BCRB}, the BCRB, and the \ac{ECRB} are explored.  
 \end{itemize}

Throughout this paper, the following notaions are used. We consider the problem of estimating the vector $\thetavec \in \Omega_\thetavec \subseteq \mathbb{R}^M$ based on the observation vector, $\xvec \in \Omega_\mathbf{x}\subseteq \mathbb{C}^N$, with joint \ac{PDF}, $f_{\mathbf{x},\thetavec}(\cdot,\cdot)$. The marginal \ac{PDF}s of $\xvec$ and $\thetavec$ are denoted by $f_\xvec(\cdot)$ and $f_\thetavec(\cdot)$, respectively, and $f_{\xvec|\thetavec}(\cdot|\cdot)$ and $f_{\thetavec|\xvec}(\cdot|\cdot)$ stand for conditional \ac{PDF}s. The expectation and the conditional expectation operators are denoted by $\text{\textnormal{E}}[\cdot]$ and $\text{\textnormal{E}}[\cdot|\cdot]$, respectively. Boldface lowercase and boldface uppercase letters are used to denote vectors and matrices, respectively. Unbold letters of either lowercase or uppercase are used for scalars. Superscripts $^T$ and $^H$ stand for transpose and conjugate transpose operations, respectively. 
The gradient of a scalar function
$b$ w.r.t. $\mathbf{a}\in\mathbb{R}^K$ is a row vector, whose $j$-th element is defined as $\left[\frac{\diff b(\mathbf{a})}{\diff \mathbf{a}}\right]_{j}\triangleq\frac{\partial b(\mathbf{c})}{\partial c_j}\Big|_{\mathbf{c}=\mathbf{a}}$. Given a vector $\mathbf{b}$, its derivative w.r.t. $\mathbf{a}$ is a matrix whose $j,k$-th entry is defined as $\left[\frac{\diff \mathbf{b}}{\diff \mathbf{a}}\right]_{j,k}\triangleq\frac{\partial b_j(\mathbf{c})}{\partial c_k}\Big|_{\mathbf{c}=\mathbf{a}}$. The notation $\mathbf{A}\succeq\mathbf{B}$ implies that $\mathbf{A}-\mathbf{B}$ is a positive-semidefinite matrix, where $\mathbf{A}$ and $\mathbf{B}$ are symmetric matrices of the same size. 
Column vectors of size $N$, whose entries are equal to zeros or ones, are denoted by $\zerovec_N$ and $\onevec_N$, respectively, and the identity matrix of size $N\times N$ is denoted by $\Imat_N$. 
$()'$ and $()''$ denote the first- and second-order derivatives, respectively. The sign $\stackrel{a}{=}$ stands for asymptotic equality. Finally, $\partial \Omega_\thetavec$ denotes the boundary of the parameter space, $\Omega_\thetavec$.

The remainder of the paper is organized as follows. Section \ref{Background} presents the BCRB and investigates its tightness conditions. The proposed class of bounds, called \ac{WBCRB}, is presented in Section \ref{proposed_bound} for the  scalar parameter case and optimized. Based on this result, a simple and asymptotically tight Bayesian bound, called \ac{AT-BCRB}, is proposed. Section \ref{Vector_Extension} presents an extension of the \ac{WBCRB} and \ac{AT-BCRB} to vector parameters. The proposed bound is evaluated in several common signal processing examples via simulations in Section \ref{Simulations}. Finally, Section \ref{Conclusion} summarizes this contribution.

\section{Background and Motivation}
\label{Background}
In this section, the BCRB is reviewed, and its tightness conditions are discussed. Specifically, the asymptotic tightness of the BCRB is compared to that of the non-Bayesian CRB (referred in the sequel as the CRB), and it will be explained why in general, the BCRB is asymptotically unattainable. This observation will be helpful for derivation of an asymptotically tight BCRB in the next section.   


\subsection{BCRB}
Let $\hat{\thetavec}(\xvec)$ be the estimator of $\thetavec$ from the observation vector $\xvec$, and $\evec = \hat{\thetavec}(\xvec) - \thetavec$ denote the corresponding estimation error. Then, using the Cauchy-Schwarz inequality, 
\begin{equation}
\label{CS}
    {\bf MSE}(\hat{\thetavec}) = \E\left[\evec \evec^T\right] \succeq \Hmat \Gmat^{-1} \Hmat^T
\end{equation}
where for an auxiliary function, $\psivec : \mathbb{C}^N \times \mathbb{R}^M \rightarrow \mathbb{R}^K$, $\Gmat = \E \left[\psivec(\xvec,\thetavec)\psivec^T(\xvec,\thetavec)\right]$ is assumed to be an invertible matrix and $\Hmat = -\E \left[\evec\psivec^T(\xvec,\thetavec)\right]$. The equality in (\ref{CS}) is satisfied iff 
\begin{equation}
\label{equality_CS}
   \hat{\thetavec}(\xvec) - \thetavec = \Cmat_B \psivec(\xvec,\thetavec)
\end{equation}
where $\Cmat_B \in \mathbb{R}^{M\times K}$ is independent of $\xvec , \thetavec$. For simplicity of notations, in the sequel, we will omit the dependency of $\psivec(\xvec,\thetavec)$ on its arguments. In general, the matrix $\Hmat$ depends on the estimator, $\hat{\thetavec}(\xvec)$. A necessary and sufficient condition to obtain an estimator-independent bound, is given by
\begin{equation}
\label{psi_condition}
\E \left[ \psivec(\xvec,\thetavec) | \xvec\right]=\zerovec_K,\;\forall \xvec\in \Omega_\xvec\;.
\end{equation}
In this case, the matrix $\Hmat$ is given by $\Hmat = \E \left[\thetavec\psivec^T(\xvec,\thetavec)\right]$ and the bound in (\ref{CS}) can be rewritten as
\begin{equation}
\label{CS2}
    {\bf MSE}(\hat{\thetavec}) \succeq \E \left[\thetavec\psivec^T\right] \E^{-1}  \left[ \psivec \psivec^T \right]\E \left[\psivec\thetavec^T\right] \;.
\end{equation}

The BCRB is obtained from (\ref{CS2}) by choosing the following auxiliary function
\begin{equation}
\psivec = \frac{\partial^T \log f_{\xvec,\thetavec}(\xvec,\thetavec)}{\partial \thetavec} \;,
\end{equation}
and let $\Jmat \define \E \left[ \Jmat_{DP}(\thetavec) \right]$ be the Bayesian FIM, where 
\begin{equation}
\label{JDP_def}
\Jmat_{DP}(\theta)\triangleq\Jmat_D(\thetavec) + \Lmat_P(\thetavec),
\end{equation}
which is composed of  the conditional FIM 
\begin{equation}
\label{JD}
\Jmat_D(\thetavec) \define \E \left[\left. \frac{\partial^T \log f_{\xvec|\thetavec}(\xvec|\thetavec)}{\partial \thetavec} \frac{\partial \log f_{\xvec|\thetavec}(\xvec|\thetavec)}{\partial \thetavec} \right| \thetavec\right] \;,
\end{equation}
and the matrix function
\begin{equation}
\label{LP}
\Lmat_P(\thetavec)\define \frac{\partial^T \log f_{\thetavec}(\thetavec)}{\partial \thetavec} \frac{\partial \log f_{\thetavec}(\thetavec)}{\partial \thetavec} \;.
\end{equation}
Then, under the assumption that $\thetavec f_{\thetavec | \xvec} (\thetavec | \xvec)=\zerovec_M, \; \forall \thetavec \in \partial \Omega_\thetavec, \; \xvec \in \Omega_\xvec$, if $\Jmat$ is invertible,
the BCRB is given by 
\begin{equation}
\label{BCRB}
{\bf MSE}(\hat{\thetavec}) \succeq \Jmat^{-1} \define {\bf BCRB} \;.
\end{equation}

\subsection{BCRB Tightness}
\label{BCRB_tightness}
The BCRB is attainable when the equality condition in (\ref{equality_CS}) is satisfied, that is 
\begin{equation}
\label{equality_BCRB1}
   \hat{\thetavec}(\xvec) - \thetavec = \Cmat_B 
   \frac{\partial^T \log f_{\xvec,\thetavec}(\xvec,\thetavec)}{\partial \thetavec} = \Cmat_B 
   \frac{\partial^T \log f_{\thetavec | \xvec}(\thetavec | \xvec)}{\partial \thetavec}\;,
\end{equation}
By considering the equality case in (\ref{BCRB}), one can verify that $\Cmat_B = \Jmat^{-1}  = {\bf BCRB}$. By integrating (\ref{equality_BCRB1}) \ac{w.r.t.} $\thetavec$, we conclude that the bound is attained iff the posterior distribution of $\thetavec$ is Gaussian:
\begin{equation}
\label{equality_BCRB2}
   \thetavec | \xvec \sim \mathcal{N}\left( \hat{\thetavec}(\xvec) , \Jmat^{-1}\right)\;.
\end{equation}
This is a restrictive condition, which may not be satisfied even asymptotically, when the number of independent and identically distributed (i.i.d.) observations goes to infinity.

If $\Jmat_{D} (\theta)$ is non-singular, the prior information is negligible in the asymptotic regime, and the \ac{ML}, \ac{MAP}, and \ac{MMSE} estimators coincide. The asymptotic \ac{MSE} of the \ac{ML} (or \ac{MAP}) estimator was shown to be \cite{tabrikian_krolik_TAP,Trees}
\begin{equation}
\label{ECRB}
{\bf MSE}(\hat{\thetavec}_{ML}) \stackrel{a}{=} \E \left [\Jmat_D^{-1}(\thetavec) \right] \triangleq {\bf ECRB} \;,
\end{equation}
where {\bf ECRB} stands for the expected CRB \cite{Trees}. Using Jensen's inequality, we obtain 
\begin{equation}
\label{ECRB_BCRB}
{\bf BCRB} \stackrel{a}{=} \E^{-1}\left[\Jmat_D(\thetavec) \right]  \preceq  \E \left [\Jmat_D^{-1}(\thetavec) \right] = {\bf ECRB} \;,
\end{equation}
where the equality holds iff $\Jmat_D(\thetavec)$ is independent of $\thetavec$. The \ac{ECRB} predicts the asymptotic \ac{MSE} of the \ac{ML} estimator, and in fact, it is a lower bound on the \ac{MSE} of any uniformly unbiased estimator, but it does not serve a valid lower bound in the Bayesian framework, where the estimators are not necessarily uniformly unbiased.

Therefore, unlike the CRB, the BCRB is asymptotically not attainable, unless the conditional FIM is independent of $\thetavec$. The reason for this discrepancy between the attainability behaviors of the CRB and BCRB can be found in their equality conditions. The equality condition of the CRB is given by
\begin{equation}
\label{equality_CRB}
   \hat{\thetavec}(\xvec) - \thetavec = \Cmat_{NB}(\thetavec) 
   \frac{\partial^T \log f_{\xvec|\thetavec}(\xvec|\thetavec)}{\partial \thetavec}\;,
\end{equation}
where $\Cmat_{NB}(\thetavec) = \Jmat_D^{-1}(\thetavec)$, and the estimator that satisfies this condition, is given by the \ac{ML}, $\hat{\thetavec} (\xvec) = \hat{\thetavec}_{ML} (\xvec)$. 

Comparing (\ref{equality_BCRB1}) with (\ref{equality_CRB}) in the asymptotic regime where the prior information is negligible (i.e. $\frac{\partial^T \log f_{\thetavec}(\thetavec)}{\partial \thetavec}$ is negligible compared to $\frac{\partial^T \log f_{\xvec|\thetavec}(\xvec|\thetavec)}{\partial \thetavec}$), it can be seen that the difference between the two conditions is that unlike $\Cmat_{NB}(\thetavec)$ in (\ref{equality_CRB}), $\Cmat_B$ in (\ref{equality_BCRB1}) is not allowed to be parameter-dependent. This is due to the fact that in the non-Bayesian framework, the expectations in the Cauchy-Schwarz inequality are w.r.t. $\xvec$, while in the Bayesian framework, they are w.r.t. both $\xvec$ and $\thetavec$. 
The \ac{ML} estimation error is asymptotically given by (\ref{equality_CRB}), which does not satisfy (\ref{equality_BCRB1}), unless  $\Jmat_D(\thetavec)$ is constant in $\thetavec$, and thus, the BCRB is not attainable. If $\Jmat_D(\thetavec)$ is constant, then (\ref{equality_CRB}) asymptotically satisfies (\ref{equality_BCRB1}) and thus, the BCRB is tight. 

The question that arises is whether there exists an auxiliary function $\psivec(\xvec,\thetavec)$, for which the condition (\ref{equality_CS}) is asymptotically satisfied. An immediate answer to this question could be that since the matrix $\Cmat_B$ should be parameter-independent, the auxiliary function would include the required $\thetavec$-dependency, $\Jmat_D^{-1}(\thetavec)$, i.e. $\psivec = \Jmat_D^{-1}(\thetavec) \frac{\partial^T \log f_{\xvec,\thetavec}(\xvec,\thetavec)}{\partial \thetavec}$. However, it can be verified that this auxiliary function does not satisfy the requirement in (\ref{psi_condition}), and thus, the bound in (\ref{CS2}) is not valid. 

In the next section, we propose an auxiliary function, which results in a valid lower bound, that is related to the BCRB. The resulting bound is found to be asymptotically tight. 


\section{The Proposed Bound}
\label{proposed_bound}
In this section, the proposed class of bounds is derived and investigated. In Subsection \ref{scalar_bound}, we derive the proposed class of bounds for a scalar parameter, and in Subsection \ref{Attainability} its attainability condition is studied. In Subsection \ref{optimization} a discrete approach for optimizing the bound is proposed. Finally, in Subsection \ref{ATBCRB_subsection} the \ac{AT-BCRB} is proposed based on the optimized \ac{WBCRB}, and its asymptotic tightness is proven. 


\subsection{Weighted BCRB}
\label{scalar_bound}
\begin{Thm}[Weighted BCRB]
\label{Weigthed_BCRB}
Let $\theta \in \Omega_\theta \subseteq \mathbb{R}$ and  $w: \Omega_\theta  \rightarrow \mathbb{R}^+$ be a differentiable function and consider the following regularity conditions,
\begin{itemize}
\item[C1)] $w(\theta) f_{\xvec,\theta}(\xvec,\theta)=0,\; \forall \theta \in \partial \Omega_\theta$, $\forall \xvec \in \Omega_\xvec$,
\item[C2)] $\theta w(\theta) f_{\xvec,\theta}(\xvec,\theta)=0,\; \forall \theta \in \partial \Omega_\theta$, $\forall \xvec \in \Omega_\xvec$,
    \item[C3)] $\left( w^2(\theta)\right)' f_{\xvec,\theta}(\xvec,\theta)=0, \; \forall \theta \in \partial \Omega_\theta$, $\forall \xvec \in \Omega_\xvec$,
\item[C4)] $ w^2(\theta) \frac{\partial f_{\xvec,\theta}(\xvec,\theta)}{\partial \theta}=0, \; \forall \theta \in \partial \Omega_\theta$, $\forall \xvec \in \Omega_\xvec$,
\end{itemize}
Then, the \ac{MSE} of any estimator of $\theta$ satisfies
\begin{equation}
\label{BCRB_inequality}
MSE(\hat{\theta}) \ge WBCRB
\end{equation}
where \ac{WBCRB} is given in the following forms.\\
1. Under Conditions C1 and C2,
\begin{align}
\label{bound_ver1}
 WBCRB =  \frac{ \E^2 \left[ w\left( \theta \right) \right] }{\E \left[ \left( w(\theta) \frac{\partial \log (f_{\xvec,\theta}(\xvec,\theta) w(\theta))}{\partial \theta} \right)^2  \right]} 
 \end{align}
2. Under Conditions, C1, C2, C3, C4,
\begin{align}
 \label{bound_ver2}
 WBCRB = -\frac{ \E^2 \left[ w\left( \theta \right) \right] }{\E \left[ w(\theta) \frac{\partial}{\partial \theta} \left( w(\theta)    \frac{\partial \log (f_{\xvec,\theta}(\xvec,\theta) w(\theta))}{\partial \theta} \right)   \right]} \;.
\end{align}

\end{Thm}
\begin{Proof}
	The proof appears in Appendix \ref{proof_scalar_bound}.
\end{Proof}
It can be verified that the two forms of the bound in Theorem \ref{Weigthed_BCRB} coincide with the BCRB if $w(\theta)  = 1$. Thus, this is a generalization of the BCRB. The first version of the bound in (\ref{bound_ver1}) is known as the BMZB, which was proposed in \cite{BMZ}. This bound has not commonly used since it is not known how to select the weighting function $w(\cdot)$ in order to tighten the bound. We will address this issue in the sequel.  The next proposition relates the bound to $J_{DP}(\theta)$  in (\ref{JDP_def}). 
\begin{Prop}[Relation to conditional Fisher information]
\label{Weighted_BCRB_FIM}
Assume the conditions C1-C3 of Theorem \ref{Weigthed_BCRB} in addition to 
\begin{itemize}
    \item [C5)] $\E \left[ \left. \frac{\partial \log f_{\xvec |\theta}(\xvec |\theta) }{\partial \theta}  \right|\theta \right]  = 0, \; \forall \theta \in \Omega_\theta$.
\end{itemize}
Then, the \ac{WBCRB} can be expressed as
{\small 
\begin{align}
\label{WBCRB_FIM}
WBCRB =&   \frac{ \E^2 \left[ w\left( \theta \right) \right] }{\E \left[ w^2\left( \theta \right) J_{DP}(\theta) \right]
  +\E \left[\left(w'\left( \theta \right) \right)^2 - \left( w^2\left( \theta \right) \right)''  \right]} \nonumber \\
  =&   \frac{ \E^2 \left[ w\left( \theta \right) \right] }{\E \left[ w^2\left( \theta \right) J_{DP}(\theta) \right]
  -\E \left[\left(w'\left( \theta \right) \right)^2 + 2 w\left( \theta \right)w''\left( \theta \right)   \right]}.
\end{align}
}where $J_{DP}(\theta)$ was defined in (\ref{JDP_def}).  
\end{Prop}
\begin{Proof}
	The proof appears in Appendix \ref{proof_Prop2}.
\end{Proof}

The BCRB can be obtained from the \ac{WWB} by choosing test-points which approach to zero. In \cite{aharon_tabrikian_WWTB}, the \ac{WWB} was extended to the case of arbitrary test-point transformation, rather than a constant shift test-point. It can be shown that the \ac{WBCRB} can be obtained from the bound in \cite{aharon_tabrikian_WWTB} with test-point transformation $\theta+w(\theta)h$ in a similar way as the relationship between the BCRB and the \ac{WWB}. Due to lack of space, the proof of this claim is omitted here.

\subsection{Attainability of the Proposed Class of Bounds} 
\label{Attainability}
In Subsection \ref{BCRB_tightness}, it was shown that the attainability condition of the BCRB is too restrictive, and unlike the CRB, its asymptotic attainability is not guaranteed. In this subsection, we investigate the equality conditions of the \ac{WBCRB} in (\ref{BCRB_inequality}), and the condition on $w(\theta)$ in order to guarantee its asymptotic attainability. 

Since the inequality of the \ac{WBCRB} in (\ref{BCRB_inequality}) is due to the Cauchy-Schwarz inequality in (\ref{CS}), the equality holds iff (\ref{equality_CS}) holds, with the corresponding auxiliary function, that is: 
\begin{equation}
\label{equality_CS_WBCRB}
   \hat{\theta}(\xvec) - \theta = w(\theta) \frac{\partial \log (f_{\xvec,\theta}(\xvec,\theta) w(\theta))}{\partial \theta} \;.
\end{equation}
The proportionality constant is omitted, since the function  $w(\cdot)$ may include it. Solution of (\ref{equality_CS_WBCRB}) for the joint \ac{PDF}, results:
\begin{align}
\label{PDF_equality_CS}
   f_{\xvec,\theta}(\xvec,\theta) & = \exp \left[ \int \frac{\hat{\theta}(\xvec) - \theta}{w(\theta)} d\theta + \eta(\xvec) - \log  w(\theta) \right] 
\end{align}
where $\eta(\xvec)$ is an arbitrary function. The corresponding posterior \ac{PDF} is
\begin{align}
\label{posterior_PDF_equality_CS}
   f_{\theta|\xvec}(\theta | \xvec) = \bar{\eta}(\xvec) \exp \left[ \hat{\theta}(\xvec) \int \frac{d\theta}{w(\theta)} - \int \frac{\theta d\theta}{w(\theta)} - \log  w(\theta) \right]
\end{align}
where $\bar{\eta}(\xvec) = \frac{\exp(\eta(\xvec)) }{f_\xvec(\xvec)}$. 
This \ac{PDF} belongs to the exponential family of distributions. It can be observed that in the case of $w(\theta) =J^{-1} = \E^{-1} \left[J_{DP}(\theta)\right]$, this condition coincides with the equality condition of the BCRB, presented in (\ref{equality_BCRB2}). The Gaussianity condition in (\ref{equality_BCRB2}) is a specific case of the exponential family in (\ref{posterior_PDF_equality_CS}), which implies that the choice of $w(\theta)$ allows to obtain a bound, which is achievable in a wider family of \ac{PDF}s. 



\begin{Prop}[Asymptotic attainability of \ac{WBCRB}]
    \label{asymptotic|_attainability}
Let $\xvec$ consist of $N_s$ observations: $\left\{\xvec_n\right\}_{n=1}^{N_s}$, which are conditionally (given $\theta$), i.i.d. Then, if $0<J_{DP}(\theta) <\infty$ and the weighting function,  satisfies $w(\theta) \stackrel{a}{=}c J_D^{-1}(\theta)$, with $0<c<\infty$, then \ac{WBCRB} is achieved by the \ac{MAP} (or \ac{ML}) estimator. 
\end{Prop}
\begin{Proof}
For i.i.d. observations, the conditional Fisher information, $J_D(\theta)$ is proportional to the number of samples, $N_s$ (see e.g. \cite{Kay}). Since $L_P(\theta)$ is finite and independent of $N_s$, then $J_{DP} \stackrel{a}{=} J_D(\theta)$. Accordingly, since $w(\theta) \stackrel{a}{=}c J_D^{-1}(\theta)$, the first term in the denominator of (\ref{WBCRB_FIM}) satisfies $\E \left[ w^2(\theta) J_{DP} (\theta)\right] \stackrel{a}{=} c^2 \E \left[ J_D^{-1}(\theta) \right] \propto \frac{1}{N_s}$.
Since $w(\theta)$ is asymptotically proportional to $\frac{1}{N_s}$, the second expectation in the denominator of (\ref{WBCRB_FIM}) is also proportional to $\frac{1}{N_s^2}$, and thus, it is asymptotically negligible compared to the first term.  Finally, substitution of the asymptotic expression of $w(\theta)$ into (\ref{WBCRB_FIM}), yields 
\begin{equation}
    WBCRB \stackrel{a}{=} 
    \frac{c^2 \E^2\left[ J_D^{-1} (\theta)\right]}{c^2\E\left[ J_D^{-1} (\theta)\right]} 
    = \E \left[ J_D^{-1} (\theta)\right] = ECRB \;,
\end{equation}
where the last equation is due to (\ref{ECRB}), and it is also equal to the asymptotic \ac{MSE} of the \ac{ML} or \ac{MAP} estimators. 
\end{Proof}

\subsection{Optimization of the \ac{WBCRB}}
\label{optimization}
In this subsection, the \ac{WBCRB} in (\ref{WBCRB_FIM}) is maximized w.r.t. the weighting function $w(\theta)$. Let $\bar{w}(\theta) \triangleq \frac{w(\theta)}{\E\left[w(\theta)\right]}$. Then, \ac{WBCRB} in (\ref{WBCRB_FIM}) can be rewritten as
\begin{equation}
\label{WBCRB_FIM3}
WBCRB =  \E^{-1} \left[ \bar{w}^2\left( \theta \right) J_{DP}(\theta) - 2\bar{w}(\theta)\bar{w}''(\theta)-(\bar{w}'(\theta))^2 \right].
\end{equation}

The bound can be maximized w.r.t. the weighting function by solving the following optimization problem
\begin{align}
\label{minimization}
& \min_{\bar{w}(\cdot)} \E \left[ \bar{w}^2\left( \theta \right) J_{DP}(\theta) - 2\bar{w}(\theta)\bar{w}''(\theta)-(\bar{w}'(\theta))^2 \right] \nonumber \\
  & s.t. \; \E\left[ \bar{w}(\theta)\right] = 1 \;.
\end{align}
This problem cannot be analytically solved. However, one can obtain a closed-form solution for its discretized version. We consider a finite  parameter space, $\Omega_\theta$, where $\theta_1,\ldots,\theta_L$ denote the uniformly sampled parameters over $\Omega_\theta$ with $\Delta$ spacing. Let $\wvec=[\bar{w}(\theta_1),\ldots,\bar{w}(\theta_L)]^T$, $\fvec= \Delta \cdot [f_\theta(\theta_1),\ldots,f_\theta(\theta_L)]^T$, $\Fmat = {\rm diag}\left(\fvec\right)$, $\Zmat={\rm diag}\left(J_{DP}(\theta_1),\ldots,J_{DP}(\theta_L)\right)$, and $\Dmat$ be a discrete-derivative matrix: 
\(D_{ij} = \frac{1}{\Delta}\left\{ \begin{array}{ll} 1 & i=j \\ 
 -1 & i=j+1 \\
 0 & {\rm otherwise} \end{array}\right.
\). 
Using these notations, vectors of first- and second-order derivatives of $w(\cdot)$ are approximately given by $\Dmat \wvec$ and $\Dmat \Dmat \wvec$, and the expectation integrals in (\ref{minimization}) can be approximated numerically, as follows:
\begin{align}
\label{expectations1}
& \E \left[ \bar{w}^2\left( \theta \right) J_{DP}(\theta) \right] \approx  \wvec^T \Zmat \Fmat \wvec, \\
\label{expectations2}
& \E \left[ \bar{w}(\theta)\bar{w}''(\theta) \right] \approx  \wvec^T \Fmat \Dmat \Dmat \wvec,  \\
\label{expectations3}
& \E \left[ (\bar{w}'(\theta))^2 \right] \approx  \wvec^T \Dmat^T \Fmat \Dmat \wvec,  \\
\label{expectations4}
&\E\left[ \bar{w}(\theta)\right] \approx  \fvec^T \wvec \;.
\end{align}
Using these terms, the objective function in (\ref{minimization}) is approximated by 
\begin{equation}
    Q(\wvec) =  \wvec^T \left( \Zmat \Fmat + \Phimat \right) \wvec \;,
    \end{equation}
where $\Phimat = - 2\Fmat \Dmat \Dmat - \Dmat^T \Fmat \Dmat$. By adding the transpose of the scalar $Q(\wvec)$ and dividing by 2, the objective function can be rewritten as
\begin{equation}
    Q(\wvec) = \wvec^T \left( \Zmat \Fmat + \bar{\Phimat} \right) \wvec \;,
\end{equation}
where 
\begin{equation}
\label{Gmat_def}
\bar{\Phimat} = \frac{\Phimat + \Phimat^T}{2}  = - \left( \Fmat \Dmat \Dmat + (\Fmat \Dmat \Dmat)^T  + \Dmat^T \Fmat \Dmat \right).
\end{equation}
The Lagrangian for the constrained minimization problem in (\ref{minimization}) is given by
\begin{align}
\label{Lagrangian}
\min_{\wvec} \left\{ \wvec^T \left( \Zmat \Fmat + \bar{\Phimat} \right) \wvec + \lambda \left(  \fvec^T \wvec  - 1\right)  \right\}.
\end{align}
The solution to this quadratic minimization problem with constraint $\fvec^T\wvec = 1$, is given by 
\begin{equation}
    \wvec_{opt} = \frac{ \left( \Zmat \Fmat + \bar{\Phimat} \right)^{-1} \fvec}{\fvec^T\left( \Zmat \Fmat + \bar{\Phimat} \right)^{-1} \fvec} .
\end{equation}
The optimal bound in this case is given by 
\begin{align}
\label{WBCRB_opt}
    WBCRB_{opt} & = \frac{1}{Q(\wvec_{opt})} =  \frac{1}{ \wvec^T_{opt} \left( \Zmat \Fmat + \bar{\Phimat} \right) \wvec_{opt} } \nonumber \\
    & =   \fvec^T\left( \Zmat \Fmat + \bar{\Phimat} \right)^{-1} \fvec .
\end{align}

It can be noted that asymptotically, when $J_{DP}(\theta)$ goes to infinity, the matrix $\bar{\Phimat}$ is negligible compared to $\Zmat \Fmat$. In this case, $WBCRB_{opt}$ is approximated by $WBCRB_{opt} \approx \fvec^T (\Zmat\Fmat)^{-1} \fvec = \onevec^T_L \Fmat \Fmat^{-1} \Zmat^{-1} \fvec = \onevec_L^T \Zmat^{-1} \fvec$, which is the discrete evaluation of $\E\left[ J_{DP}^{-1}(\theta) \right] \stackrel{a}{=} \E\left[ J_{D}^{-1}(\theta) \right] =ECRB$, which indicates that the bound is asymptotically attanable. 

Equation (\ref{WBCRB_opt}) provides an expression for discrete evaluation of the optimal \ac{WBCRB}. However, it involves sampling the parameter space, and inversion of a matrix of size $L$, which represents the amount of samples. Except the fact that the method may suffer from high computational complexity when the number of samples is large, the lower bound property may be lost, because it is an approximated solution. In the following, we assume that the number of samples over the parameter space, $L$, approaches infinity, such that the expression in (\ref{WBCRB_opt}) is accurate. After some manipulations which slightly loosen the bound, we will obtain a simple and asymptotically tight lower bound.   Loosening of the bound is based on an inequality given in the following lemma. 
\begin{Lemm}
\label{Lemma_inequality}
  If $(\Imat_L + \Psimat) \succ 0$, then $(\Imat_L + \Psimat)^{-1} \succeq (\Imat_L - \Psimat)$.
\end{Lemm}
\begin{Proof}
    Let $\Umat$ and $\Lambdamat={\rm diag}(\lambda_1,\ldots,\lambda_L)$ denote the eigenvectors and eigenvalues matrices of $\Psimat$. Then, 
    \begin{equation}
        (\Imat_L + \Psimat)^{-1} = (\Umat\Umat^T + \Umat\Lambdamat \Umat^T)^{-1} =\Umat ( \Imat_L + \Lambdamat )^{-1} \Umat^T .
    \end{equation}
Since $1+\lambda_l>0$, it can be verified that $\frac{1}{1+\lambda_l} \ge 1-\lambda_l, \;\forall l=1,\ldots,L$, then $(\Imat_L + \Lambdamat)^{-1} \succeq \Imat -\Lambdamat$. Multiplying the left and right hand sides by $\Umat$ and $\Umat^T$, respectively, results in $(\Imat_L + \Psimat)^{-1} \succeq (\Imat_L - \Psimat)$.
    \end{Proof}

The optimal bound in (\ref{WBCRB_opt}) can be rewritten as 
\begin{align}
\label{WBCRB_opt2}
     WBCRB_{opt}=  \hspace{6.2cm} \nonumber \\
     \fvec^T(\Zmat \Fmat)^{-1/2} \left( \Imat_L + (\Zmat \Fmat)^{-1/2}\bar{\Phimat} (\Zmat \Fmat)^{-1/2}\right)^{-1} (\Zmat \Fmat)^{-1/2}\fvec \nonumber \\ 
      \ge \fvec^T(\Zmat \Fmat)^{-1/2} \left( \Imat_L - (\Zmat \Fmat)^{-1/2}\bar{\Phimat} (\Zmat \Fmat)^{-1/2}\right) (\Zmat \Fmat)^{-1/2}\fvec \nonumber \\ 
      =\fvec^T(\Zmat \Fmat)^{-1}\fvec - \fvec^T(\Zmat \Fmat)^{-1}\bar{\Phimat} (\Zmat \Fmat)^{-1}\fvec , \hspace{2.55cm} \nonumber \\
      =\onevec_L^T\Zmat^{-1}\fvec - \onevec_L^T\Zmat^{-1}\bar{\Phimat} \Zmat^{-1}\onevec , \hspace{4cm}
\end{align}
where the inequality is due to Lemma \ref{Lemma_inequality}, and in the last equality we used $\Fmat \onevec_L = \fvec$. Substitution of (\ref{Gmat_def}) into  (\ref{WBCRB_opt2}) yields
\begin{align}
\label{WBCRB_opt3}
WBCRB_{opt}  \ge   \onevec_L^T\Zmat^{-1}\fvec & + \onevec_L^T\Zmat^{-1} \Fmat \Dmat \Dmat \Zmat^{-1}\onevec_L \nonumber \\ 
&  + \onevec_L^T\Zmat^{-1} (\Fmat \Dmat \Dmat)^T \Zmat^{-1}\onevec_L  \nonumber \\  & + \onevec_L^T\Zmat^{-1} \Dmat^T \Fmat \Dmat \Zmat^{-1}\onevec_L  .  
\end{align}
The terms in the right hand side (r.h.s.) of (\ref{WBCRB_opt3}) are discrete evaluations of the following expectations:
\begin{align}
\label{expectations_backward}
&  \onevec_L^T\Zmat^{-1}\fvec = \E \left[ J_{DP}^{-1}(\theta)\right],  \\ 
& \onevec_L^T\Zmat^{-1} \Dmat^T \Fmat \Dmat \Zmat^{-1}\onevec_L = \E \left[ ((J_{DP}^{-1}(\theta))')^2 \right], \\
& \onevec_L^T\Zmat^{-1} (\Fmat \Dmat \Dmat)^T \Zmat^{-1}\onevec_L 
 =  \onevec_L^T\Zmat^{-1} \Fmat \Dmat \Dmat \Zmat^{-1}\onevec_L \nonumber \\
 & = \E \left[ J_{DP}^{-1}(\theta) \left( J_{DP}^{-1}(\theta) \right)''\right]  .  
\end{align}
In terms of expectations, (\ref{WBCRB_opt3}) can be rewritten as
\begin{align}
\label{WBCRB_opt4}
WBCRB_{opt}  \ge  WBCRB_{sub}    
\end{align}
where $WBCRB_{sub}$ is a suboptimal bound defined as
\begin{align}
\label{WBCRB_opt4_def}
WBCRB_{sub} \triangleq & \E \left[ J_{DP}^{-1}(\theta)\right] + \E \left[ ((J_{DP}^{-1})')^2 \right]  \nonumber \\ 
& + 2 \E \left[ J_{DP}^{-1}(\theta) \left( J_{DP}^{-1}(\theta) \right)''\right].  
\end{align}
Any function $q(\theta)$, which is twice-differentiable, satisfies: $2qq'' + (q')^2=(q^2)''-(q')^2$. Using this identity, we obtain
\begin{align}
\label{WBCRB_opt5}
 WBCRB_{sub} = \E \left[ J_{DP}^{-1}(\theta)\right] + \E \left[ (J_{DP}^{-2})'' - ((J_{DP}^{-1}(\theta))')^2 \right].  
\end{align}
In the asymptotic regime, the second term in the r.h.s. of (\ref{WBCRB_opt5}) is negligible compared to the first term, which asymptotically equals the \ac{ECRB}. Equation (\ref{WBCRB_opt5}) provides a valid lower bound that is asymptotically tight. 

\subsection{\ac{AT-BCRB}}
\label{ATBCRB_subsection}
The next theorem proposes a simple version of the \ac{WBCRB}, called \ac{AT-BCRB}. In the next section, we will show that this bound is asymptotically tight and always tighter than the suboptimal \ac{WBCRB} from (\ref{WBCRB_opt5}). 

\begin{Thm}[AT-BCRB]
\label{AT-BCRB_Thm}
If $J_{DP}(\theta)$ is invertible and twice differentiable $\forall \theta \in \Omega_\theta$, then the \ac{MSE} of any estimator of $\theta$ satisfies
 \begin{equation}
 \label{ATBCRB}
MSE(\hat{\theta}) \ge ATBCRB 
\end{equation}
where
 \begin{equation}
 \label{ATBCRB_def}
ATBCRB \triangleq \frac{ \E \left[ J_{DP}^{-1} (\theta) \right] }{1 + \rho}
\end{equation}
and 
\begin{align}
\label{rho}
\rho \triangleq  \frac{\E \left[\left( \left(J_{DP}^{-1}(\theta) \right)'\right)^2 - \left( J_{DP}^{-2}(\theta) \right)''  \right] }{\E \left[ J_{DP}^{-1}(\theta) \right] }.
\end{align}


\end{Thm}
\begin{Proof}
	By setting $w(\theta) = J_{DP}^{-1}(\theta)$, and its substitution into \ac{WBCRB} in (\ref{WBCRB_FIM}), one obtains
\begin{align}
\label{ATBCRB_exact}
     MSE(\hat{\theta}) \ge WBCRB|_{w(\theta) = J_{DP}^{-1}(\theta)} \hspace{3.3cm}& \nonumber \\
     =  \frac{ \E^2 \left[ J_{DP}^{-1}(\theta) \right] }{\E \left[ J_{DP}^{-1}(\theta) \right]
  +\E \left[\left( \left(J_{DP}^{-1}(\theta) \right)'\right)^2 - \left( J_{DP}^{-2}(\theta) \right)''  \right]} \nonumber \\ 
  = \frac{ \E \left[ J_{DP}^{-1}(\theta) \right] }{1
  + \rho}.
\end{align}

\end{Proof}


\begin{Prop}
\label{ATBRB_ECRB}(Asymptotic relationship between \ac{AT-BCRB} and \ac{ECRB})
Let $\xvec$ consist of $N_s$ observations: $\left\{\xvec_n\right\}_{n=1}^{N_s}$, which  are assumed to be conditionally (given $\theta$) i.i.d. Then, if $L_P(\theta)<\infty, \forall \theta \in \Omega_\theta$, the \ac{AT-BCRB} asymptotically coincides with the \ac{ECRB}: 
        \begin{equation}
        \label{ATBCRB_ECRB_relation}
            \lim_{N\rightarrow \infty } \frac{ATBCRB}{ECRB} = 1.
        \end{equation}
\end{Prop}
\begin{Proof}
For (conditionally) i.i.d. observations, $J_D(\theta) \propto N_s$. Thus, according to the definition of $J_{DP}(\theta)$, for bounded $L_P(\theta)$, $ J_{DP}^{-1}(\theta) \xrightarrow{\:  N_s \rightarrow \infty \: } J_D^{-1}(\theta) \propto 1/N_s$. The derivatives of $J_{DP}^{-1}(\theta)$ do not change the asymptotic dependency on $N$, and thus, by observing (\ref{rho}), one obtains $\lim_{N_s\rightarrow \infty} \rho = 0$. Substitution of these asymptotic terms into (\ref{ATBCRB_def}) and the definition of \ac{ECRB} in (\ref{ECRB}) leads to (\ref{ATBCRB_ECRB_relation}). 
\end{Proof}
As mentioned above, the asymptotic performances of the \ac{ML} or \ac{MAP} estimators are given by the \ac{ECRB} \cite{tabrikian_krolik_TAP,Trees}, however, it is not a lower bound. Proposition \ref{ATBRB_ECRB} implies that the \ac{AT-BCRB}, which is a lower bound on any estimator, expresses the asymptotic performance of the \ac{ML} or \ac{MAP} estimators.

\begin{Prop}[Order relation between \ac{AT-BCRB} and the suboptimal \ac{WBCRB}]
    \label{ATBCRB_vs_Suboptimal}
 $ATBCRB \ge WBCRB_{sub}$.
\end{Prop}
\begin{Proof}
Using the definition of \ac{AT-BCRB} in (\ref{ATBCRB_def}) and the inequality $\frac{1}{1+\rho}\ge 1-\rho$ for any $\rho$, which satisfies $1+\rho \ge 0$, one obtains
\begin{align}
        ATBCRB & =  \frac{ \E \left[ J_{DP}^{-1} (\theta) \right] }{1 + \rho} \nonumber \\
        & \ge \E \left[ J_{DP}^{-1} (\theta) \right] (1 - \rho) = WBCRB_{sub} \;.
\end{align}
where the last equation is due to the definition of $\rho$ in (\ref{rho}) and $WBCRB_{sub}$ in (\ref{WBCRB_opt5}). 
\end{Proof}

Propositions \ref{ATBRB_ECRB} and  \ref{ATBCRB_vs_Suboptimal} indicate that the \ac{AT-BCRB} provides a lower bound that is a good approximation of the optimal \ac{WBCRB}, and it is asymptotically attainable, while it appears in a simple form.  

\section{Extension to Vector Parameters}
\label{Vector_Extension}

In this section, the \ac{WBCRB} and the \ac{AT-BCRB}, proposed in the previous section for scalar parameters, are extended to the case of vector parameters. 

\begin{Thm}[Weighted BCRB for vector parameter]
\label{Weigthed_Matrix_BCRB}
Let $\thetavec \in \Omega_\thetavec \subseteq \mathbb{R}^M$ and $\Wmat: \Omega_\thetavec  \rightarrow \mathbb{R}^{M\times M}$ be a symmetric, positive-definite matrix whose entries are differentiable functions. 
Then, under the following regularity conditions,
\begin{itemize}
\item[C6)] $\Wmat(\thetavec) f_{\xvec,\thetavec}(\xvec,\thetavec)=\zerovec_{M\times M},\; \forall \thetavec \in \partial \Omega_\thetavec$, $\forall \xvec \in \Omega_\xvec$,
\item[C7)] $\Wmat(\thetavec) \onevec_M \thetavec^T f_{\xvec,\thetavec}(\xvec,\thetavec)=\zerovec_{M\times M},\; \forall \thetavec \in \partial \Omega_\thetavec$, $\forall \xvec \in \Omega_\xvec$,
\item[C8)] $W_{mn}(\thetavec) \frac{\partial W_{ik}(\thetavec)}{\partial \theta_k} f_{\thetavec} (\thetavec) =0, \; \forall \thetavec \in \partial \Omega_\thetavec, \; n,m,i,k=1,\ldots,M$, 
\item[C9)] $\E \left[ \left. \frac{\partial^T\log f_{\xvec|\thetavec} (\xvec|\thetavec)}{\partial \thetavec}    \right|\thetavec \right] =\zerovec_M, \;\; \forall \thetavec \in \Omega_\thetavec $,
\end{itemize}
the \ac{MSE} matrix of any estimator of $\thetavec$ satisfies:
\begin{equation}
\label{BCRB_matrix_inequality}
{\bf MSE}(\hat{\thetavec}) \succeq {\bf WBCRB} = \E \left[ \Wmat(\thetavec) \right] \Fmat^{-1} \E \left[ \Wmat(\thetavec) \right]
\end{equation}
where
{\small{
\begin{align} 
\label{Fmat}
\Fmat = &  
 \E \left[ \Wmat(\thetavec) \Jmat_{DP}(\thetavec) \Wmat(\thetavec) \right]  
- \E \left[ {\rm div}\left(\Wmat(\thetavec)\right) {\rm div}^T\left(\Wmat(\thetavec)\right) \right] 
 \nonumber \\ &
 - \E\left[ \Wmat(\thetavec) \frac{\partial^T}{\partial \thetavec} {\rm div}\left(\Wmat(\thetavec)\right) \right] 
 - \E\left[  \frac{\partial}{\partial \thetavec} {\rm div}\left(\Wmat(\thetavec)\right) \Wmat(\thetavec) \right]
\end{align}
}}
\hspace{-.1cm}and the vector ${\rm div}(\Wmat(\thetavec))$ denotes the divergence of the matrix $\Wmat(\thetavec)$, whose $m$th element is given by
\begin{equation}
    \left[{\rm div}\left(\Wmat(\thetavec)\right)\right]_m = \sum_{n=1}^M \frac{\partial W_{mn}(\thetavec)}{\partial \theta_n} 
\end{equation}
\end{Thm}
\begin{Proof}
    The proof appears in Appendix \ref{Matrix_Proof}.
\end{Proof}

The matrix bound in (\ref{BCRB_matrix_inequality}) provides an extension of the scalar bound in (\ref{WBCRB_FIM}) for parameter vector estimation. The results of the previous section provide an indication of how to choose the weighting matrix in order to obtain a simple and asymptotically tight bound. Thus, assuming that $\Jmat_{DP}^{-1}(\thetavec)$ is invertible for $\thetavec \in \Omega_\thetavec$, we propose the following weighting matrix $\Wmat(\thetavec) = \Jmat_{DP}^{-1}(\thetavec)$. In this case, (\ref{BCRB_matrix_inequality}) and (\ref{Fmat}) can be rewritten as
\begin{equation}
\label{BCRB_matrix_inequality_AT}
{\bf MSE}(\hat{\thetavec}) \succeq {\bf ATBCRB} = \E \left[ \Jmat_{DP}^{-1}(\thetavec) \right] \Fmat^{-1} \E \left[ \Jmat_{DP}^{-1}(\thetavec) \right]
\end{equation}
where
{\small{
\begin{align} 
\label{Fmat_AT}
\Fmat = &  
 \E \left[ \Jmat_{DP}^{-1}(\thetavec) \right]  
- \E \left[ \dvec(\thetavec) \dvec^T(\thetavec) \right] 
 \nonumber \\ &
 - \E\left[ \Jmat_{DP}^{-1}(\thetavec) \frac{\partial^T \dvec(\thetavec)}{\partial \thetavec} \right] 
 - \E\left[  \frac{\partial \dvec(\thetavec)}{\partial \thetavec}  \Jmat_{DP}^{-1}(\thetavec) \right]
\end{align}
}}
and $\dvec(\thetavec) = {\rm div}\left( \Jmat_{DP}^{-1}(\thetavec) \right)$.

\section{Simulation Results}
\label{Simulations}
In this section, we evaluate the proposed bounds in three signal processing examples, and demonstrate their tightness compared to the \ac{BCRB} and the \ac{ECRB}. First, we consider the problem of estimating the variance of a zero-mean white Gaussian random process. Then, we  address the problem of \ac{DOA} estimation using a sensor array. Finally, we consider a vector parameter case for estimation of the mean of Gaussian random variable with unknown variance in the presence of prior statistical information. 

\subsection{Variance Estimation}
Consider the variance estimation problem of a zero-mean Gaussian-distributed random variable using conditionally \ac{iid} samples. The observation vector consists of $N$ samples, $\xvec=\left[x_1, \dots ,x_N \right]^T$, satisfying $\xvec|\theta \sim  \mathcal{N}\left(\zerovec_N, \theta\Imat_N\right)$, where $\theta$ 
 is the unknown parameter. We assume the beta prior distribution for $\theta$:
\begin{equation}
\label{eq:betaPDF}
    f_\theta(\theta)=\frac{1}{\beta(a,b)}\theta^{a-1}(1-\theta)^{b-1}, \;\;\;\; \theta \in \Omega_\theta = [0,1]
\end{equation}
where $\beta(a,b)$ is a beta function, defined as
\begin{equation}
\label{eq:betaFun}
    \beta(a,b) \triangleq \int_0^1\theta^{a-1}(1-\theta)^{b-1}d\theta=\frac{\Gamma(a)\Gamma(b)}{\Gamma(a+b)},
\end{equation}
and $\Gamma(a)\define\int_0^\infty v^{a-1}e^{-v}dv$ is the gamma function. 
    The parameters $a$ and $b$ are real, positive values that determine the shape of the distribution. Hereafter, we assume $a=b$ and consequentially, the \ac{PDF} in (\ref{eq:betaPDF}) is symmetric w.r.t. its mean, $\mu_\theta=\frac{1}{2}$, and the variance is $\sigma_\theta^2=\frac{1}{4(2a+1)}$.
When $a=1$ the prior \ac{PDF} is uniform and as $a$ increases, the \ac{PDF} becomes narrower.
The regularity conditions in the BCRB are satisfied  when $a>2$. 

The \ac{AT-BCRB} was computed using (\ref{ATBCRB_def}), where the conditional Fisher information is $J_D(\theta)=\frac{N}{2\theta^2}$, and using (\ref{LP}), $L_P(\theta)=(a-1)^2\left( \frac{1}{\theta}-\frac{1}{1-\theta} \right)^2$. The derivatives of $w(\theta)=J_{DP}^{-1}(\theta)$ were analytically computed, whereas the expectations in \ac{AT-BCRB} were evaluated via numerical integration. The optimal bound, was computed using (\ref{WBCRB_opt}), where the sampling step was set to $\Delta=0.02$.

The BCRB and the \ac{ECRB}, as well as  \ac{MAP} and \ac{ML} estimators for this problem were derived in \cite[pages 7-11]{Trees}. The performances of the two estimators were evaluated using 5000 Monte-Carlo trials, where for each realization, the variance, $\theta$ was randomized based on the prior distribution described above, and then the data samples where randomized using the conditional distribution $\xvec|\theta$.  

\begin{figure}[htb]
\centering 
\includegraphics[width=0.48\textwidth]{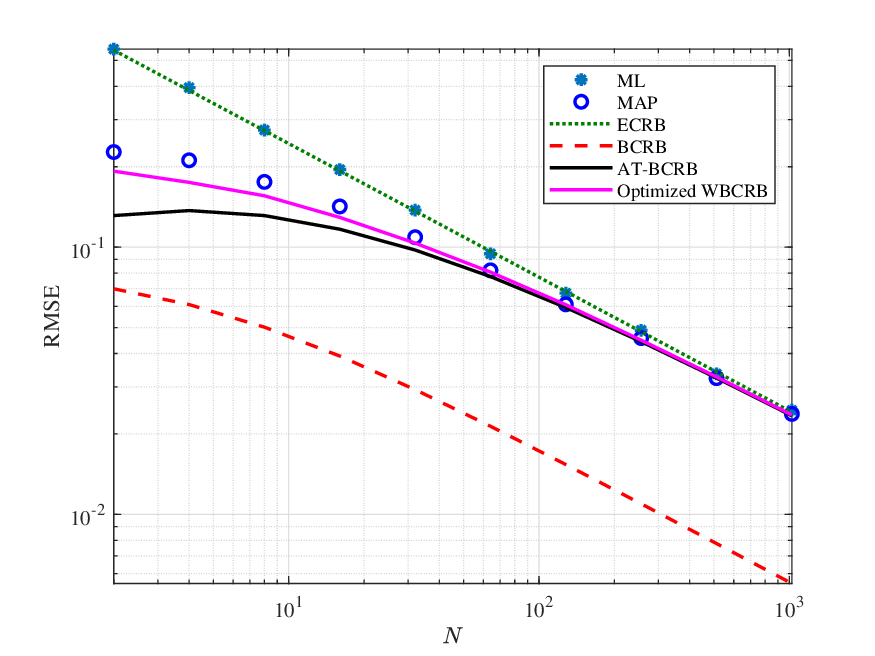}
\caption{The proposed bounds compared to BCRB, \ac{ECRB}, and the \ac{RMSE}s of \ac{MAP} and \ac{ML} for variance estimation of conditionally white Gaussian samples, as a function of the number of samples under prior beta distribution with shape parameters $a=b=2.1$.}
\label{fig:var_est}
\end{figure}

Fig. \ref{fig:var_est} presents the \ac{AT-BCRB} and the optimized \ac{WBCRB}, compared to \ac{BCRB}, \ac{ECRB}, and the \ac{RMSE} of \ac{MAP} and \ac{ML} estimators, as a function of the number of samples, $N$. The prior \ac{PDF} shape parameter, $a$, was set to $a=2.1$. It can be observed that the \ac{AT-BCRB} and the optimal \ac{WBCRB} are asymptotically attained by the \ac{MAP} and \ac{ML} 
estimators. Also, the \ac{ECRB} provides a good indication of the asymptotic performance of the two estimators, but it can be seen that it does not provide a valid lower bound, since the \ac{RMSE} of the \ac{MAP} estimator is below the \ac{ECRB}. As mentioned above, the ECRB serves as a  lower bound (on the expected MSE) for uniformly unbiased estimators. Thus, in this problem, it can serve as a lower bound for the ML estimator since it is uniformly unbiased. The MAP estimator, which  is biased in this problem, is not bounded by the ECRB. The optimized \ac{WBCRB} provides a good prediction of the \ac{MAP} performance, also at low number of samples. It can be noticed that the BCRB that is commonly used in the signal processing community, is far from the \ac{RMSE}s of the estimators, and it is not tight {\it even asymptotically}. 

\begin{figure}[htb]
\centering 
\includegraphics[width=0.48\textwidth]{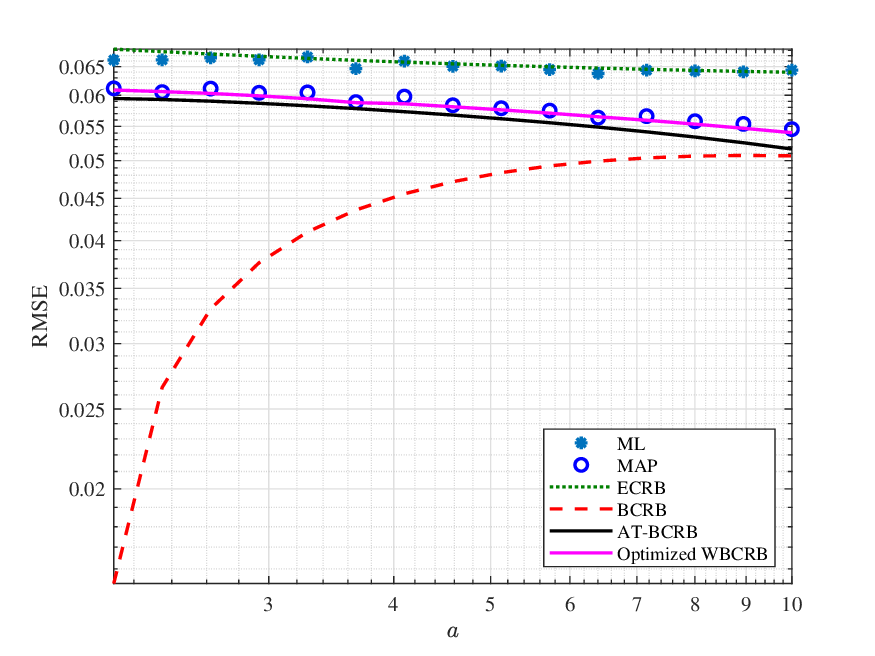}
\caption{The proposed bounds compared to BCRB, \ac{ECRB}, and the \ac{RMSE}s of \ac{MAP} and \ac{ML} for variance estimation of conditionally white Gaussian samples, under prior beta distribution as a function of the shape parameter, $a$ with $b=a$ and $N=128$ samples.}
\label{fig:var_LBvsA}
\end{figure}

Fig. \ref{fig:var_LBvsA} depicts the bounds and the estimators as a function of the prior \ac{PDF} shape parameter, $a$, with $N=128$ samples. It can be seen that also here, the \ac{AT-BCRB} provides a good prediction of the performance of the \ac{MAP} estimator, while the optimal \ac{WBCRB} coincides with \ac{MAP} \ac{RMSE} for all the tested values of the shape parameter, $a$. The tightness of the BCRB is degraded when $a$ approaches 2, where the prior \ac{PDF} becomes closer to uniform distribution and thus, the probability of obtaining small values of $\theta$ increases. At small values of $a$, the dependency of the conditional Fisher information $J_D(\theta)$ on $\theta$ is stronger. On the other hand, when $a$ increases, the probability for obtaining small values of $\theta$ decreases, and the average dependency of $J_D(\theta)$ on $\theta$ is weaker. In the extreme case of constant $J_D(\theta)$,  
Jensen's inequality in (\ref{ECRB_BCRB}) becomes an equality. Therefore, the gap between the BCRB and the \ac{ECRB} is expected to reduce when the average dependency of $J_D(\theta)$ on $\theta$ becomes smaller.

\subsection{DOA Estimation}
\label{sec:DOA}
In this example, we address the problem of \ac{DOA} estimation of a narrowband far-filed source using a uniform linear array of $N_a$ sensors with half a wavelength spacing. The data model in this problem is given by 
\begin{equation}
\label{eq:DOA}
\xvec_j=\alpha_j\avec(\theta)+\vvec_j, \;\; j=1,\dots,N_s
\end{equation}
 where $\theta\in (-85,85)\frac{\pi}{180}$ is the source DOA and $\avec(\theta)$ is the steering vector whose elements are given by 
 $a_n(\theta) =e^{i\pi \left(n-\frac{N_a-1}{2} \right) \sin \theta},\;\;n=0,\dots,N_a-1$.  The signal and noise sequences, $\left\{\alpha_j\right\}$ and 
 $\left\{\vvec_j\right\}$, are white Gaussian processes with distributions: $\alpha_j \sim \mathcal{N}^C (0,\sigma_\alpha^2)$ and $\vvec_j\sim \mathcal{N}^C(\zerovec_{N_a}, \sigma^2 \Imat_{N_a})$. The random terms in the model, $\theta$, $\{\alpha_j\}$, and $\{\vvec_j\}$, are assumed to be statistically mutually independent. 
The signal and noise variances, $\sigma_\alpha^2$ and $\sigma^2$, are assumed to be known. 

Since the proposed bounds, as well as the BCRB, do not exist for uniform prior distribution, we use a smooth version of it based on a normalized raised-cosine filter as follows:
\begin{equation}
        \label{RCF_PDF}
            f_{\theta}(\theta)=\frac{1}{C}\left\{ 
   \begin{array}{ll}
   1, &  \left|\theta\right|<s \kappa \\
   \frac{1}{2}\left( 1+\cos \frac{\pi \left( \left| \theta \right|-s\kappa \right) }{s(1-\kappa)} \right),
   & s\kappa \leqslant \left| \theta \right| \leqslant s\\
   0, &  \left| \theta \right| > s
   \end{array} 
\right.
\end{equation}
where $C$ is a normalizion factor, $s$ defines the filter's edges and $\kappa \in [0,1]$ controls the edges slope (roll-off factor). For example, $\kappa=1$ and $\kappa=0$ result in uniform and pure raised cosine filters, respectively. We set $s=85\frac{\pi}{180}$ and $\kappa=0.98$ to approximate a uniform distribution in $\left( -s,s \right)$. 

It can be verified that the conditional Fisher information for estimating $\theta$ is given by
\begin{equation}
    J_D(\theta)=\frac{2N_s N_a  SNR^2}{1+N_a SNR}\left\|\dot{\avec}(\theta)\right\|^2=\delta\cos^2\theta
    \label{eq:DOA_JD}
\end{equation}
where SNR is defined as $SNR\triangleq\frac{\sigma_\alpha^2}{\sigma^2}$,  $\delta \triangleq \frac{2\pi^2 N_sN_a SNR^2}{1+N_a \cdot SNR} \times$ $\sum_{n = 0}^{N_a - 1} \left(n - \frac{N_a-1}{2}\right)^2 =  \frac{\pi^2 N_s SNR^2}{1+N_a \cdot SNR} \frac{N_a^2(N_a^2-1)}{6}$, and $L_P(\theta)=\left( \frac{\partial \log f_{\theta}(\theta)}{\partial \theta} \right)^2$.
The derivatives of the AT-BCRB weighting function, $w(\theta)=J_{DP}^{-1}(\theta) =\left(J_{D}(\theta) + L_{P}(\theta)\right)^{-1}$, were analytically computed, whereas the expectations \ac{w.r.t.} $\theta$ for computing the bounds were evaluated via numerical integration.

In the example, we assumed $N_a=32$ sensors, and evaluated the following bounds: AT-BCRB in (\ref{ATBCRB_def}), optimal WBCRB in (\ref{WBCRB_opt}), BCRB in (\ref{BCRB}), ECRB in (\ref{ECRB}), and the \ac{RMSE} of the \ac{MAP} estimator for the DOA, $\theta$. The performance of the MAP estimator was evaluated using 5000 Monte-Carlo trials. The performance of the ML estimator in this example is very close to the MAP estimator, since the prior distribution is almost uniform.  

\begin{figure}[htb]
\centering 
\includegraphics[width=0.48\textwidth]{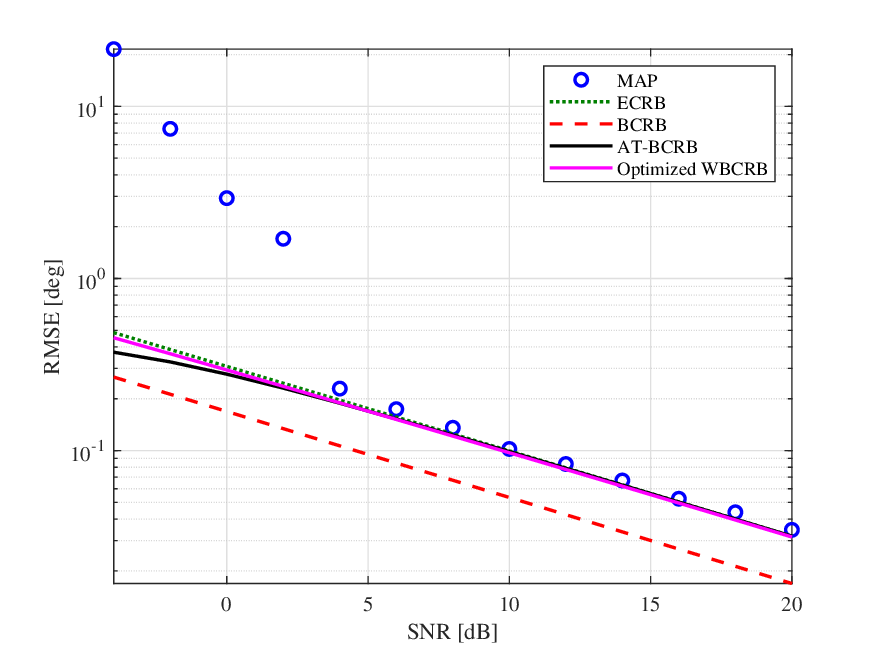}
\caption{The compared bounds and the MAP estimator RMSE for DOA estimation versus SNR, with $N_a=32$ sensors and $N_s=128$ snapshots.}
\label{fig:DOA_SNR}
\end{figure}

\begin{figure}[htb]
\centering 
\includegraphics[width=0.48\textwidth]{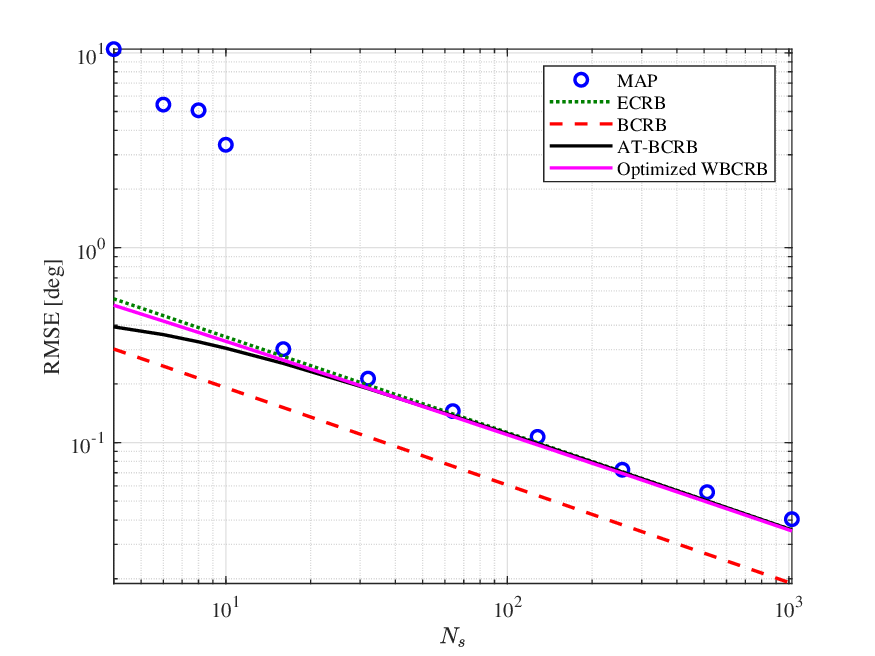}
\caption{The compared bounds and the MAP estimator RMSE for DOA estimation versus the number of snapshots, with $N_a=32$ sensors and SNR=10dB.}
\label{fig:DOA_J}
\end{figure}

Figs. \ref{fig:DOA_SNR} and \ref{fig:DOA_J} compare the bounds and the RMSE of the MAP estimator versus SNR with $N_s=128$ snapshots, and versus the number of snapshots $N_s$ with $SNR=10$dB, respectively. The figures show that the BCRB does not reliably predict the performance of the MAP estimator, even at high SNR region or large number of snapshots, while the AT-BCRB, the optimal \ac{WBCRB}, and the \ac{ECRB} coincide with the \ac{RMSE} of the \ac{MAP} estimator above the threshold SNR. It can be seen in both figures that the AT-BCRB is very close to the optimal \ac{WBCRB} above the threshold region. In this example, the ECRB serves as a lower bound, because above the threshold region, the MAP estimator is nearly uniformly unbiased, however, in general, the ECRB is not guaranteed to lower bound the performance as seen in the previous example.  

\subsection{Mean Estimation of a Gaussian Process with Unknown Variance in the Presence of Prior Statistical Information}
In this subsection, we consider the problem of mean estimation of a (conditionally) white Gaussian random process with prior statistical information on the mean and variance. This example illustrates implementation of the matrix version of the AT-BCRB. Let $\xvec=\left[x_1, \dots ,x_N \right]^T$ denote the observation vector and the unknown parameter vector is $\thetavec=[\mu \; \varphi]^T$, such that $\xvec|\thetavec \sim  \mathcal{N}\left(\mu\onevec_N, \varphi\Imat_N\right)$. We assume that $\mu$ and $\varphi$ are statistically independent random variables with prior distributions: $\mu \sim \mathcal{N}(0,\sigma_{\mu}^2)$ and $\varphi \sim \text{Beta}(a,b)$ as defined in (\ref{eq:betaPDF}), with $a=b$.

The conditional FIM for estimating $\thetavec$ is given by 
\begin{equation}
\label{J_D_MAT}
    \Jmat_D(\thetavec) = \left[\begin{matrix}
 \frac{N}{\varphi} & 0  \\
 0 & \frac{N}{2\varphi^2}
\end{matrix}\right]
\end{equation}
and the matrix $\Lmat_P(\thetavec)$ can be obtained using (\ref{LP}):
\begin{equation}
\label{L_P_MAT}
    \Lmat_{P}(\thetavec) = \left[\begin{matrix}
  \left(\frac{\mu}{\sigma_{\mu}^2} \right)^2 & \frac{\mu}{\sigma_{\mu}^2}\gamma(\varphi)  \\
 \frac{\mu}{\sigma_{\mu}^2}\gamma(\varphi) &  \gamma^2(\varphi)
\end{matrix}\right],
\end{equation}
where $\gamma(\varphi) \define \left(a-1\right)\left(\frac{1}{1-\varphi} - \frac{1}{\varphi}\right)$. The AT-BCRB was computed using (\ref{BCRB_matrix_inequality_AT}) and (\ref{Fmat_AT}), where the required partial derivatives in (\ref{Fmat_AT}) were computed numerically. The \ac{BCRB} and the \ac{ECRB} were analytically computed using (\ref{BCRB}) and (\ref{ECRB}), respectively:
\begin{align}
\label{BCRB_MAT}
    {\bf BCRB}=&\E^{-1}\left[ \Jmat_{DP} \right] \nonumber \\=&
    \left[\begin{matrix}
     \left( \frac{1}{\sigma_{\mu}^2} + \frac{N(2a-1)}{a-1} \right)^{-1} & 0  \\
    0 & \frac{a-2}{(2a-1)\left( N+4(a-1) \right)}
\end{matrix}\right]\\
\label{ECRB_MAT}
    {\bf ECRB}=&\E\left[ \Jmat_{D}^{-1} \right] =
    \left[\begin{matrix}
    \frac{1}{2N} & 0  \\
     0 & \frac{a+1}{N(2a+1)}
\end{matrix}\right] .
\end{align}
The shape parameter was set to $a=2.1$. The performance of the \ac{MAP} estimator was evaluated using 5000 Monte-Carlo trials.  

\begin{figure}[htb]
\centering 
\includegraphics[width=0.48\textwidth]{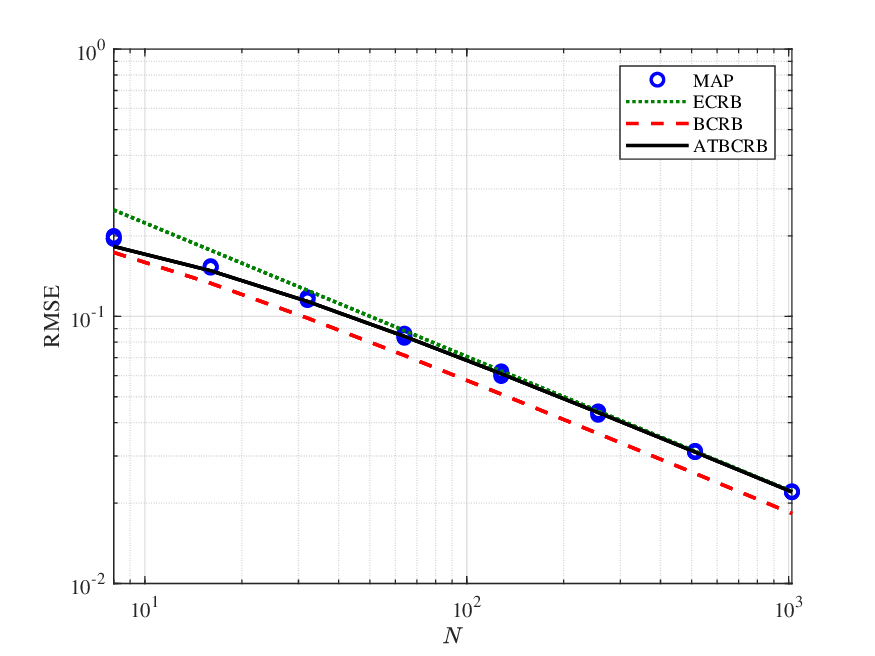}
\caption{AT-BCRB, BCRB, ECRB, and the RMSE of the MAP estimator for the mean of a Gaussian process, $\mu$, with unknown variance, $\varphi$, using $N$ conditionally \ac{iid} observations with available prior statistical information:  $\mu \sim \mathcal{N}(0,\sigma_{\mu}^2)$ with $\sigma_{\mu}^2=0.1$ and $\varphi \sim \text{Beta}(a,b)$ with $a=b=2.1$.}
\label{fig:RMSE_Vs_N}
\end{figure}

Fig. \ref{fig:RMSE_Vs_N} presents the \ac{AT-BCRB} compared to the \ac{BCRB}, the \ac{ECRB}, and the \ac{RMSE} of the \ac{MAP} estimator as a function of the number of samples, $N$, with  $\sigma_{\mu}^2=0.1$. The figure shows that the AT-BCRB is close to the performance of the \ac{MAP} estimator and is asymptotically achieved by it. In contrast, the BCRB is not tight even asymptotically. Although the \ac{ECRB} predicts the asymptotic performance of the \ac{MAP} estimator, it does not provide a valid lower bound at low number of samples.

\begin{figure}[htb]
\centering 
\includegraphics[width=0.48\textwidth]{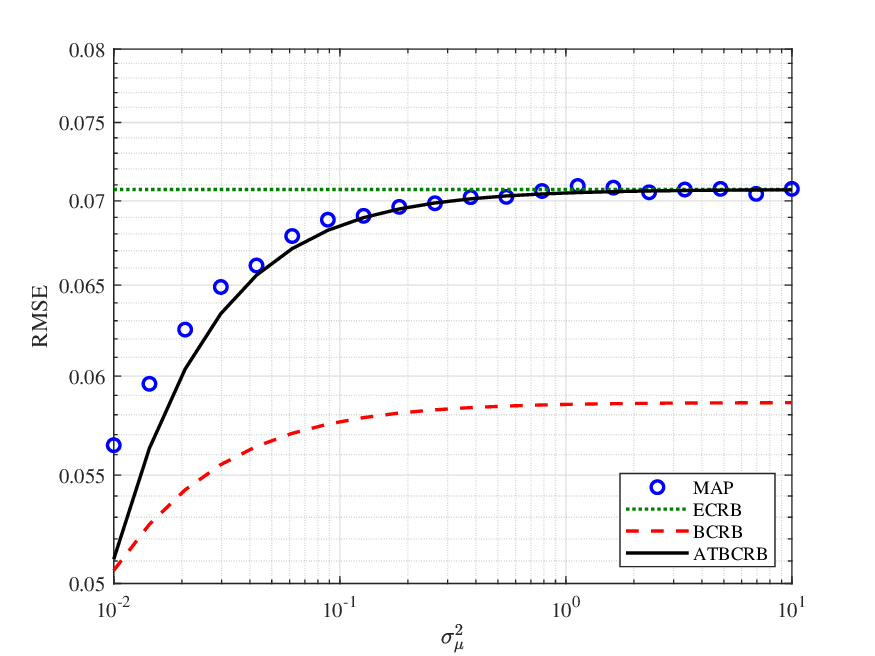}
\caption{AT-BCRB, BCRB, ECRB, and the RMSE of the MAP estimator for the mean of a Gaussian process, $\mu$, with unknown variance, $\varphi$, using $N=100$ conditional \ac{iid} observations with available prior statistical information: $\mu \sim \mathcal{N}(0,\sigma_{\mu}^2)$ and $\varphi \sim \text{Beta}(a,b)$ with $a=b=2.1$.}
\label{fig:RMSE_Vs_mu}
\end{figure}

Fig. \ref{fig:RMSE_Vs_mu} depicts the bounds \ac{AT-BCRB}, \ac{BCRB}, \ac{ECRB}, and the RMSE of the \ac{MAP} estimator versus $\sigma_{\mu}^2$ for $N=100$ samples. The figure shows that also here the AT-BCRB provides a lower bound that reliably predicts the performance of the MAP estimator. However, the BCRB is not achievable for all the tested values of $\sigma_{\mu}^2$. For large prior variance, $\sigma_{\mu}^2$, the estimator relies mainly on the observations, and therefore, its bias vanishes. Accordingly, the ECRB, which assumes uniform unbiasedness, is tight in this region. However, for small values of $\sigma_{\mu}^2$, where the \ac{MAP} estimator is biased, the ECRB does not lower bound it. 

\section{Conclusion}
\label{Conclusion}
This paper addresses the problem of \ac{MSE} lower bound in the Bayesian framework. Unlike the non-Bayesian CRB, which is known to be asymptotically tight, the BCRB is not attainable, even asymptotically, unless the conditional FIM is independent of the parameter. This paper solves this problem and proposes a Bayesian Cram\'{e}r-Rao-type bound, which is asymptotically attainable. First, a class of Cram\'{e}r-Rao type bounds in the Bayesian framework is proposed. This class is related to the weighted BCRB and is extended to vector parameters. The bound in its scalar form is maximized w.r.t. the weighting function, and it is shown that the optimal weighting function is related to the conditional Fisher information. Based on this result, a new simple bound, called \ac{AT-BCRB} is proposed for both scalar and vector parameter, and it is shown that it asymptotically coincides with the \ac{ECRB}, which expresses the asymptotic performance of the \ac{ML} or \ac{MAP} estimators. Unlike the \ac{ECRB}, the \ac{AT-BCRB} is a valid lower bound in Bayesian parameter estimation.  

Based on the results of this paper, the signal processing theory and methods which involve the BCRB, can be revisited by using the proposed simple bound. These applications include cases where the bound is used for estimators' performance benchmark, system design tool,  or criterion for cognitive systems, where a performance measure is required as a function of system parameters. 

\appendices
\section{Proof of Theorem 1}
\label{proof_scalar_bound}
\begin{Proof}
The proof is based on the inequality in (\ref{CS2}) using the following auxiliary function:
\begin{equation}
\label{psi_g}
    \psivec = w(\theta) \frac{\partial \log (f_{\xvec,\theta}(\xvec,\theta) w(\theta))}{\partial \theta}\;.
\end{equation}
We will first show that this auxiliary function satisfies the condition in (\ref{psi_condition}), and then prove the two versions of the bound as stated in the theorem. 

By substitution of (\ref{psi_g}) into (\ref{psi_condition}), we obtain
\begin{align}
\label{psi_condition_with_g}
\E \left[ \psi | \xvec\right]= & \E \left[ \left. w(\theta) \frac{\partial \log (f_{\xvec,\theta}(\xvec,\theta) w(\theta))}{\partial \theta} \right| \xvec\right] \\
= & 
\E \left[ \left. \left( w(\theta) \frac{\partial \log (f_{\theta|\xvec}(\theta|\xvec) )}{\partial \theta} + w'(\theta) \right) \right| \xvec\right] \\
= & \int_{\Omega_\theta} \left( w(\varphi) \frac{\partial f_{\theta|\xvec}(\varphi|\xvec) }{\partial \varphi} + w'(\varphi) f_{\theta|\xvec}(\varphi|\xvec) \right) d \varphi \\
= & \int_{\Omega_\theta}  \frac{\partial} {\partial \varphi} \left( w(\varphi) f_{\theta|\xvec}(\varphi|\xvec) \right) d \varphi =0,\;\forall \xvec\in \Omega_\xvec \;,
\end{align}
where the last equality is due to condition C1 of the theorem.

For computation of the bound in (\ref{CS2}), the terms $\E \left[\theta\psi\right]$ and  $\E  \left[ \psi^2 \right]$ are required. Using the auxiliary function, $\psi$, in (\ref{psi_g}), yields
{\small{
\begin{align}
\label{E_theta_psi}
\E \left[ \theta \psi \right]= & \E \left[ \theta  w(\theta) \frac{\partial \log (f_{\xvec,\theta}(\xvec,\theta) w(\theta))}{\partial \theta} \right] \\
= & \int_{\Omega_\xvec} \int_{\Omega_\theta} \varphi w(\varphi) \frac{\partial \log (f_{\xvec,\theta}(\xivec,\varphi) w(\varphi) )}{\partial \varphi} f_{\xvec,\theta}(\xivec,\varphi) d \varphi d \xivec\\
= & \int_{\Omega_\xvec}\int_{\Omega_\theta} \varphi \left( w(\varphi) \frac{\partial f_{\xvec,\theta}(\xivec,\varphi)}{\partial \varphi}  +  w'(\varphi) f_{\xvec,\theta}(\xivec,\varphi) \right) d \varphi d \xivec \\
= & \int_{\Omega_\xvec}\int_{\Omega_\theta} \varphi \frac{\partial} {\partial \varphi} \left( w(\varphi) f_{\xvec,\theta}(\xvec,\varphi) \right) d \varphi d\xivec \;.
\label{E_theta_psi2}
\end{align}
}}
\hspace{-0.1cm}By integration by parts of (\ref{E_theta_psi2}) and using Condition C2 of the theorem, we obtain
\begin{equation}
\label{E_theta_psi3}
\E \left[ \theta \psi \right]= - \int_{\Omega_\xvec} \int_{\Omega_\theta} w(\varphi) f_{\xvec,\theta}(\xvec,\varphi) d \varphi d\xivec
= - \E \left[  w(\theta) \right]  
 \;.
\end{equation}
From the definition of $\psi$ in (\ref{psi_g}), we obtain
\begin{align}
\label{E_psi_squared}
\E \left[ \psi^2 \right]= \E \left[ \left(  w(\theta) \frac{\partial \log (f_{\xvec,\theta}(\xvec,\theta) w(\theta))}{\partial \theta} \right)^2 \right] \;.
\end{align}
Substitution of (\ref{E_theta_psi3}) and  (\ref{E_psi_squared}) into (\ref{CS2}) results in the bound in (\ref{bound_ver1}).

The second version of the bound can be obtained as follows:
{\small{
\begin{align}
\label{E_psi_squared3}
\E \left[ \psi^2 \right]=  \hspace{7.5cm} \nonumber \\ =\int_{\Omega_\xvec} \int_{\Omega_\theta}  \left(  w(\varphi) \frac{\partial \log (f_{\xvec,\theta}(\xivec,\varphi) w(\varphi))}{\partial \varphi} \right)^2 f_{\xvec,\theta}(\xivec,\varphi) d\varphi d\xivec  \hspace{0.8cm} \nonumber \\
= \int_{\Omega_\xvec} \int_{\Omega_\theta}  \frac{\partial (f_{\xvec,\theta}(\xivec,\varphi) w(\varphi))}{\partial \varphi} w(\varphi)    \frac{\partial \log (f_{\xvec,\theta}(\xivec,\varphi) w(\varphi))}{\partial \varphi} d\varphi d\xivec
\;.
\end{align}
}}
Applying integration by parts on (\ref{E_psi_squared3}), results
{\small{
\begin{align}
\label{E_psi_squared4}
\E \left[ \psi^2 \right]=& \int_{\Omega_\xvec} \left. f_{\xvec,\theta}(\xivec,\varphi) w^2(\varphi)   \frac{\partial \log (f_{\xvec,\theta}(\xivec,\varphi) w(\varphi))}{\partial \varphi} \right|_{\partial \Omega_\theta}  d\xivec \nonumber \\ 
& - \int_{\Omega_\xvec} \int_{\Omega_\theta}  f_{\xvec,\theta}(\xivec,\varphi) w(\varphi)   \times \nonumber \\ 
& \frac{\partial}{\partial \varphi} \left( w(\varphi) \frac{\partial \log (f_{\xvec,\theta}(\xivec,\varphi) w(\varphi))}{\partial \varphi} \right)  d\varphi d\xivec \\
=& \int_{\Omega_\xvec} \left.  w(\varphi)   \frac{\partial 
 (f_{\xvec,\theta}(\xivec,\varphi) w(\varphi))}{\partial \varphi} \right|_{\partial \Omega_\theta}  d\xivec \nonumber \\ 
 \label{E_psi_squared5}
& - \E \left[w(\theta) \frac{\partial}{\partial \theta} \left( w(\theta)    \frac{\partial \log (f_{\xvec,\theta}(\xivec,\theta) w(\theta))}{\partial \theta} \right) \right]
\;.
\end{align}
}}
\hspace{-0.1cm}Under conditions C3 and C4, the first term in the r.h.s. of (\ref{E_psi_squared5}) is equal to zero. Finally, by substitution of (\ref{E_theta_psi3}) and (\ref{E_psi_squared5}) into (\ref{CS2}) one obtains the second form of the bound in (\ref{bound_ver2}).

\end{Proof}

\section{Proof of Proposition \ref{Weighted_BCRB_FIM}}
\label{proof_Prop2}
\begin{Proof}
Let us consider the denominator of (\ref{bound_ver1}):
\begin{align}
\label{bound_ver1_denom}
& \E \left[ \left( w(\theta) \frac{\partial \log (f_{\xvec,\theta}(\xvec,\theta) w(\theta))}{\partial \theta} \right)^2  \right]  \nonumber \\ 
& = 
\E \left[ \left( w(\theta) \frac{\partial \log f_{\xvec,\theta}(\xvec,\theta) }{\partial \theta} + w'(\theta)\right)^2  \right] \nonumber  \\ 
& = 
\E \left[ w^2(\theta) \left( \frac{\partial \log f_{\xvec,\theta}(\xvec,\theta) }{\partial \theta}\right)^2  \right] + \E \left[ \left(w'(\theta)\right)^2  \right] \nonumber \\ 
&   \hspace{3cm} + 2\E  \left[ w(\theta) w'(\theta) \frac{\partial \log f_{\xvec,\theta}(\xvec,\theta) }{\partial \theta} \right] \;,
\end{align}
The last term in the r.h.s. of (\ref{bound_ver1_denom}) can be derived using integration by parts, as follows:
\begin{align}
\label{gdgdL}
&  2\E \left[ w(\theta) w'(\theta) \frac{\partial \log f_{\xvec,\theta}(\xvec,\theta) }{\partial \theta} \right]  \nonumber \\  
&  = \int_{\Omega_\xvec} \int_{\Omega_\theta}  \left( w^2(\varphi)\right)' \frac{\partial f_{\xvec,\theta}(\xivec,\varphi) }{\partial \varphi} d \varphi d \xivec \nonumber \\ 
&  =  \int_{\Omega_\xvec} \left.  \left( w^2(\varphi)\right)' f_{\xvec,\theta}(\xivec,\varphi)\right|_{\partial \Omega_\theta} d \xivec \nonumber \\ 
&  \hspace{0.4cm} - \int_{\Omega_\xvec} \int_{\Omega_\theta}  \left( w^2(\varphi)\right)'' f_{\xvec,\theta}(\xivec,\varphi) d \varphi d \xivec \nonumber \\
&  = -\E \left[ \left( w^2(\theta) \right)'' \right] \;.
\end{align}

The first and second terms in the r.h.s. of (\ref{bound_ver1_denom}) can be derived using law of total expectation, as follows:
\begin{align}
\label{Eg2L2}
    & \E \left[ w^2(\theta) \left( \frac{\partial \log f_{\xvec,\theta}(\xvec,\theta) }{\partial \theta}\right)^2  \right]  \nonumber \\ 
    & = \E  \left[  w^2(\theta) \E \left[ \left. \left( \frac{\partial \log f_{\xvec |\theta}(\xvec |\theta) }{\partial \theta} + \frac{\partial \log f_{\theta}(\theta) }{\partial \theta} \right)^2 \right|\theta \right] \right] \nonumber \\
    & = \E  \left[  w^2(\theta) \E \left[ \left. \left( \frac{\partial \log f_{\xvec |\theta}(\xvec |\theta) }{\partial \theta} \right)^2 \right|\theta \right] \right] \nonumber \\ 
    & + 2\E  \left[  w^2(\theta) \frac{\partial \log f_{\theta}(\theta) }{\partial \theta}  \E \left[ \left. \frac{\partial \log f_{\xvec |\theta}(\xvec |\theta) }{\partial \theta}  \right|\theta \right] \right] \nonumber \\ 
    & +  \E  \left[  w^2(\theta)  \left( \frac{\partial \log f_{\theta}(\theta) }{\partial \theta} \right)^2  \right] \nonumber \\ 
    & = \E  \left[  w^2(\theta)  J_D(\theta)  \right] + \E  \left[  w^2(\theta)  L_P(\theta)  \right] = \E \left[  w^2(\theta)  J_{DP}(\theta)  \right] 
\end{align}
where the third equality is obtained by using the definitions in (\ref{JD}) and (\ref{LP}), and the fact that the mixed term is equal to zero, due to Condition C5 of the proposition. 

Substitution of (\ref{gdgdL}) and (\ref{Eg2L2}) into (\ref{bound_ver1_denom}) yields 
\begin{align}
\label{bound_ver1_denom_final}
& \E \left[ \left( w(\theta) \frac{\partial \log (f_{\xvec,\theta}(\xvec,\theta) w(\theta))}{\partial \theta} \right)^2  \right]  \nonumber \\ 
& = \E \left[  w^2(\theta)  J_{DP}(\theta)  \right] + \E \left[ \left(w'(\theta)\right)^2 - \left( w^2(\theta) \right)'' \right]  \;.
\end{align}
Finally, by substitution of (\ref{bound_ver1_denom_final}) into the denominator of (\ref{bound_ver1}), we obtain the bound in (\ref{WBCRB_FIM}).
\end{Proof}

\section{Proof of Theorem \ref{Weigthed_Matrix_BCRB}}
\label{Matrix_Proof}
The proof follows the lines of the proof of the bound for the scalar case in Appendix \ref{proof_scalar_bound}. The derivation is based on the matrix inequality in (\ref{CS2}). After introducing the auxiliary function, $\psivec(\xvec,\thetavec)$, the condition in (\ref{psi_condition}) will be examined, and the terms in (\ref{CS2}) will be derived. 

Consider the following auxiliary function:
\begin{equation}
    \label{auxiliary_vector}
    \psivec(\xvec,\thetavec) = \Wmat(\thetavec) \frac{\partial^T\log f_{\xvec,\thetavec} (\xvec,\thetavec)}{\partial \thetavec} + {\rm div}\left(\Wmat(\thetavec)\right) .
\end{equation}
The $i$-th element of $\psivec(\xvec,\thetavec)$ is given by
\begin{align}
    \label{psi_i}
    \psi_i(\xvec,\thetavec) & = \sum_{n=1}^M \left( W_{in}(\thetavec) \frac{\partial\log f_{\xvec,\thetavec} (\xvec,\thetavec)}{\partial \theta_n} + \frac{\partial W_{in}(\thetavec)}{\partial \theta_n} \right)  \\
    \label{psi_i_2}
    & = \sum_{n=1}^M  W_{in}(\thetavec) \frac{\partial\log \left( f_{\xvec,\thetavec} (\xvec,\thetavec) W_{in}(\thetavec) \right) }{\partial \theta_n} .
 \end{align}
By substitution of (\ref{psi_i}) into (\ref{psi_condition}), we obtain
\begin{align}
\label{psi_condition_vector}
\E \left[ \psi_i | \xvec\right]  \hspace{7.7cm} \nonumber \\ 
= \sum_{n=1}^M
\E \left[ \left. \left( W_{in}(\thetavec) \frac{\partial \log (f_{\thetavec|\xvec}(\thetavec|\xvec) )}{\partial \theta_n} + \frac{\partial W_{in}(\thetavec)}{\partial \theta_n} \right) \right| \xvec\right] \hspace{1cm} \\
= \sum_{n=1}^M \int_{\Omega_\thetavec} \left( W_{in}(\varphivec) \frac{\partial f_{\thetavec|\xvec}(\varphivec|\xvec) }{\partial \varphi_n} + \frac{\partial W_{in}(\varphivec)}{\partial \varphi_n} f_{\thetavec|\xvec}(\varphivec|\xvec) \right) d \varphivec \hspace{0.1cm}\\
= \sum_{n=1}^M \int_{\Omega_\thetavec}  \frac{\partial} {\partial \varphi_n} \left( W_{in}(\varphivec) f_{\thetavec|\xvec}(\varphivec|\xvec) \right) d \varphivec =0,\;\forall \xvec\in \Omega_\xvec \;, \hspace{0.7cm}
\end{align}
where the last equality is due to Condition C6 of the theorem. Accordingly, the condition in (\ref{psi_condition}) is satisfied, and thus the matrix inequality in (\ref{CS2}) can be used. 

Now, we derive the matrices $\E \left[\thetavec \psivec^T\right]$ and $\E \left[\psivec \psivec^T\right]$  from (\ref{CS2}). The $mi$-th element of $\E \left[\thetavec \psivec^T\right]$ is given by
\begin{align}
    \label{E_theta_psi_mi}
    \left[ \E \left[ \thetavec \psivec^T\right] \right]_{mi} = \E \left[\theta_m\psi_i \right] \hspace{5.1cm} \nonumber \\
    =\sum_{n=1}^M \E \left[ \theta_m  \left( W_{in}(\thetavec) \frac{\partial\log f_{\xvec,\thetavec} (\xvec,\thetavec)}{\partial \theta_n} + \frac{\partial W_{in}(\thetavec)}{\partial \theta_n} \right)
    \right] \hspace{1cm}\nonumber \\
=\sum_{n=1}^M\int_{\Omega_\xvec}\int_{\Omega_\theta} \varphi_m \left( W_{in}(\varphivec) \frac{\partial f_{\xvec,\thetavec}(\xivec,\varphivec)}{\partial \varphi_n}  \right. \hspace{3cm} \nonumber \\ 
\left. + \frac{\partial W_{in}(\varphivec )}{\partial \varphi_n} f_{\xvec,\thetavec}(\xivec,\varphivec) \right) d \varphivec d \xivec \nonumber \\
=\sum_{n=1}^M \int_{\Omega_\xvec}\int_{\Omega_\theta} \varphi_m \frac{\partial} {\partial \varphi_n} \left( W_{in}(\varphivec) f_{\xvec,\theta}(\xvec,\varphivec) \right) d \varphivec d\xivec . \hspace{1.4cm}
\end{align}
Integration by parts of the r.h.s. of (\ref{E_theta_psi_mi}) yields
\begin{align}
    \label{E_theta_psi_mi2}
    &\left[ \E \left[ \thetavec \psivec^T\right] \right]_{mi} = \nonumber \\ 
    &= \sum_{n=1}^M \int_{\Omega_\xvec} \int_{\Omega_\thetavec \backslash \Omega_{\theta_n} }\left. \varphi_m 
      W_{in}(\varphivec) f_{\xvec,\theta}(\xvec,\varphivec) \right|_{\partial \Omega_{\theta_n}} d\varphivec^{(n)} d\xivec  \hspace{-1cm} \nonumber \\
  &\hspace{0.4cm} -\sum_{n=1}^M \int_{\Omega_\xvec}\int_{\Omega_\theta} \delta_{mn}  W_{in}(\varphivec) f_{\xvec,\theta}(\xvec,\varphivec) d \varphivec d\xivec \hspace{2.2cm}  \nonumber \\
 &=  - \int_{\Omega_\xvec}\int_{\Omega_\theta} W_{im}(\varphivec) f_{\xvec,\theta}(\xvec,\varphivec) d \varphivec d\xivec   = \E \left[ W_{im}(\varphivec)\right]\hspace{-1cm} 
\end{align}
where $d\varphivec^{(n)}$ is the differential of the vector $\varphivec$ after excluding $\varphi_n$, $\delta_{mn}$ is the Kronecker delta, and the second equality is obtained under condition C7 of the theorem. Accordingly, 
\begin{equation}
\label{E_theta_PSI}
    \E \left[ \thetavec \psivec^T\right] = \E \left[ \Wmat^T(\thetavec) \right] =  \E \left[ \Wmat(\thetavec) \right] \;.
\end{equation}

Using (\ref{auxiliary_vector}),  $\E \left[\psivec \psivec^T\right]$ is given by
\begin{align} 
\label{E_psi_psi_T}
\E \left[ \psivec \psivec^T \right]  
=  \E \left[ \left( \Wmat(\thetavec) \frac{\partial^T\log f_{\xvec,\thetavec} (\xvec,\thetavec)}{\partial \thetavec} + {\rm div}\left(\Wmat(\thetavec)\right) \right) \right. \nonumber \\ 
 \hspace{2cm} \times \left. \left( \Wmat(\thetavec) \frac{\partial^T\log f_{\thetavec} (\thetavec)}{\partial \thetavec} + {\rm div}\left(\Wmat(\thetavec)\right) \right)^T
\right] 
\nonumber \\ 
= \Smat_1 + \Smat_2 + \Smat_2^T + \Smat_3 \hspace{3.8cm}
\end{align}
where
\begin{align}
\label{S1}
\Smat_1 \triangleq  & \E \left[  \Wmat(\thetavec) \frac{\partial^T\log f_{\xvec,\thetavec} (\xvec,\thetavec)}{\partial \thetavec}  \frac{\partial \log f_{\xvec,\thetavec} (\xvec,\thetavec)}{\partial \thetavec} \Wmat(\thetavec) \right] 
 \\ 
\label{S2}
  \Smat_2 \define & \E \left[ \Wmat(\thetavec) \frac{\partial^T\log f_{\xvec,\thetavec} (\xvec,\thetavec)}{\partial \thetavec}  {\rm div}^T\left(\Wmat(\thetavec)\right)  \right] 
 \\ 
 \label{S3}
  \Smat_3 \define & \E \left[ {\rm div} \left(\Wmat(\thetavec)\right)  {\rm div}^T\left(\Wmat(\thetavec)\right)   \right]. 
\end{align}
Using the law of total expectation, $\Smat_1$ from (\ref{S1}) can be simplified as follows: 
\begin{align} 
\label{S1_final}
\Smat_1 = &\E \left[  \Wmat(\thetavec) \frac{\partial^T\log f_{\xvec,\thetavec} (\xvec,\thetavec)}{\partial \thetavec}  \frac{\partial \log f_{\xvec,\thetavec} (\xvec,\thetavec)}{\partial \thetavec} \Wmat(\thetavec) \right]  
\nonumber \\
= & \E \left[ \Wmat(\thetavec) \Jmat_{DP}(\thetavec) \Wmat(\thetavec) \right] 
\end{align}
where 
\begin{align} 
\label{J_DP_mat}
\Jmat_{DP}(\thetavec) \define & \E \left[ \left. \frac{\partial^T\log f_{\xvec,\thetavec} (\xvec,\thetavec)}{\partial \thetavec}  \frac{\partial \log f_{\xvec,\thetavec} (\xvec,\thetavec)}{\partial \thetavec}  \right|\thetavec \right]  \nonumber \\ 
= & \E \left[ \left. \frac{\partial^T\log f_{\xvec|\thetavec} (\xvec|\thetavec)}{\partial \thetavec}  \frac{\partial \log f_{\xvec|\thetavec} (\xvec|\thetavec)}{\partial \thetavec}  \right|\thetavec \right]  \nonumber \\ 
& + \E \left[ \left. \frac{\partial^T\log f_{\xvec|\thetavec} (\xvec|\thetavec)}{\partial \thetavec}    \right|\thetavec \right] \frac{\partial \log f_{\thetavec} (\thetavec)}{\partial \thetavec}  \nonumber \\ 
& + \frac{\partial^T \log f_{\thetavec} (\thetavec)}{\partial \thetavec}  \E \left[ \left. \frac{\partial\log f_{\xvec|\thetavec} (\xvec|\thetavec)}{\partial \thetavec}  
  \right|\thetavec \right]  \nonumber \\ 
& + \frac{\partial^T \log f_{\thetavec} (\thetavec)}{\partial \thetavec} \frac{\partial\log f_{\xvec|\thetavec} (\xvec|\thetavec)}{\partial \thetavec} \nonumber \\ 
= & \Jmat_{D}(\thetavec) + \Lmat_P(\thetavec)
\end{align}
and $\Jmat_{D}(\thetavec)\define \E \left[ \left. \frac{\partial^T\log f_{\xvec|\thetavec} (\xvec|\thetavec)}{\partial \thetavec}  \frac{\partial \log f_{\xvec|\thetavec} (\xvec|\thetavec)}{\partial \thetavec}  \right|\thetavec \right]$, $\Lmat_P(\thetavec) \define \frac{\partial^T \log f_{\thetavec} (\thetavec)}{\partial \thetavec} \frac{\partial\log f_{\thetavec} (\thetavec)}{\partial \thetavec}$. In (\ref{J_DP_mat}), the mixed terms vanish,  according to Condition C9 of the theorem. 

The term $\Smat_2$ from (\ref{S2}) can be written as:
\begin{align} 
\label{E_psi_psi_T_term2}
\Smat_2 = &  \E \left[  \Wmat(\thetavec) \E \left[ \left.  \frac{\partial^T\log f_{\xvec,\thetavec} (\xvec,\thetavec)}{\partial \thetavec}   \right| \thetavec \right]  {\rm div}^T\left(\Wmat(\thetavec)\right) \right] \nonumber \\ 
= &\E \left[  \Wmat(\thetavec)   \frac{\partial^T\log f_{\thetavec} (\thetavec)}{\partial \thetavec}  {\rm div}^T\left(\Wmat(\thetavec)\right) \right], 
\end{align}
where the first equality is due to law of total expectation and in the second equality we used  Condition C9 of the theorem. 

The $mi$-th element of (\ref{E_psi_psi_T_term2}) can be expressed as
\begin{align} 
\label{E_psi_psi_T_term2_elements}
\left[\Smat_2\right]_{mi} = & \sum_{n=1}^M  \E \left[  W_{mn}(\thetavec)  \frac{\partial\log f_{\thetavec} (\thetavec)}{\partial \theta_n}  \left[ {\rm div}\left(\Wmat(\thetavec)\right) \right]_i \right] \nonumber \\ 
= &\sum_{n=1}^M \int_{\Omega_{\thetavec}} W_{mn}(\varphivec) \left[ {\rm div}\left(\Wmat(\varphivec)\right) \right]_i \frac{\partial f_{\thetavec} (\varphivec)}{\partial \varphi_n} d\varphivec \;.
\end{align}
The multivariate integral in (\ref{E_psi_psi_T_term2_elements}) is over  $\Omega_{\theta_1},\ldots,\Omega_{\theta_M}$. Using integration by parts, the integral over $\Omega_{\theta_n}$ is given by
\begin{align}
\label{int_temp1}
\int_{\Omega_{\theta_n}} W_{mn}(\varphivec)  \left[ {\rm div}\left(\Wmat(\varphivec)\right) \right]_i \frac{\partial f_{\thetavec} (\varphivec)}{\partial \varphi_n} d\varphi_n  \hspace{2.4cm}
\nonumber \\ 
 = \left. W_{mn}(\varphivec)  \left[ {\rm div}\left(\Wmat(\varphivec)\right) \right]_i f_{\thetavec} (\varphivec)\right|_{\partial \Omega_{\varphi_n}} \hspace{2.7cm} \nonumber \\ 
 - \int_{\Omega_{\theta_n}} \frac{\partial }{\partial \varphi_n } \left(W_{mn}(\varphivec)  \left[ {\rm div}\left(\Wmat(\varphivec)\right) \right]_i\right) f_{\thetavec} (\varphivec)  d\varphi_n 
 \end{align}
Under Condition C8 of the theorem, the first term in the r.h.s. of (\ref{int_temp1}) is equal to zero. Thus, substitution of (\ref{int_temp1}) into the r.h.s. of (\ref{E_psi_psi_T_term2_elements}) results in 
{\small{
\begin{align}
\left[\Smat_2\right]_{mi} = & - \sum_{n=1}^M \int_{\Omega_{\thetavec}} \frac{\partial }{\partial \varphi_n } \left(W_{mn}(\varphivec)  \left[ {\rm div}\left(\Wmat(\varphivec)\right) \right]_i \right) f_{\thetavec} (\varphivec)  d\varphivec \nonumber \\
=& - \sum_{n=1}^M \E \left[ \frac{\partial }{\partial \theta_n } \left(W_{mn}(\thetavec)  \left[ {\rm div}\left(\Wmat(\thetavec)\right) \right]_i \right) \right]  \nonumber \\
= & -\E \left[ \sum_{n=1}^M \frac{\partial W_{mn}(\thetavec)}{\partial \theta_n } \left( \left[ {\rm div}\left(\Wmat(\thetavec)\right) \right]_i \right) \right]  \nonumber \\ 
 & - \E \left[ \sum_{n=1}^M W_{mn}(\thetavec) \frac{\partial}{\partial \theta_n } \left[ {\rm div}\left(\Wmat(\thetavec)\right) \right]_i  \right]  \nonumber \\
 =& -\E \left[ \left[ {\rm div}\left(\Wmat(\thetavec)\right) \right]_m  \left[ {\rm div}\left(\Wmat(\thetavec)\right) \right]_i \right]  \nonumber \\
 & - \E \left[ \sum_{n=1}^M W_{mn}(\thetavec) \left[ \frac{\partial^T}{\partial \thetavec} {\rm div}\left(\Wmat(\thetavec)\right) \right]_{ni}  \right] \;.
\end{align}
}}
Accordingly, the matrix $\Smat_2$ can be expressed as
{\small{
\begin{equation}
\label{S2_final}
\Smat_2 = \hspace{-1mm}- \E \left[ {\rm div}\left(\Wmat(\thetavec)\right) {\rm div}^T\left(\Wmat(\thetavec)\right) \right] - \E\left[ \Wmat(\thetavec) \frac{\partial^T}{\partial \thetavec} {\rm div}\left(\Wmat(\thetavec)\right) \right].
\end{equation}
}}
Substitution of the terms $\Smat_1$, $\Smat_2$, and $\Smat_3$ from (\ref{S1_final}), (\ref{S2_final}), and (\ref{S3}) into (\ref{E_psi_psi_T}) yields 
\begin{align} 
\label{E_psi_psi_T_final}
\E \left[ \psivec \psivec^T \right]  
= &  \E \left[ \Wmat(\thetavec) \Jmat_{DP}(\thetavec) \Wmat(\thetavec) \right]  
 \nonumber \\ &
- \E \left[ {\rm div}\left(\Wmat(\thetavec)\right) {\rm div}^T\left(\Wmat(\thetavec)\right) \right] 
 \nonumber \\ &
 - \E\left[ \Wmat(\thetavec) \frac{\partial^T}{\partial \thetavec} {\rm div}\left(\Wmat(\thetavec)\right) \right] 
\nonumber \\  & 
 - \E\left[  \frac{\partial}{\partial \thetavec} {\rm div}\left(\Wmat(\thetavec)\right) \Wmat(\thetavec) \right]\;.
\end{align}
By substituting (\ref{E_theta_PSI}) and (\ref{E_psi_psi_T_final}) into (\ref{CS2}), we obtain the bound in (\ref{BCRB_matrix_inequality}).

\bibliographystyle{IEEEtran}


\end{document}